# Curl-based Electric-Field Boundary Condition for the Accurate and Stable Electromagnetic Scattering Analysis

Eduard Ubeda, Alex Heldring, Juan M. Rius and Vladimir Okhmatovski

***Abstract*— We introduce a curl-based Electric-Field Integral Equation (Curl-EFIE) for the electromagnetic scattering analysis from perfect electric conductors. The formulation is derived by enforcing a vanishing curl on the EFIE over the boundary manifold, achieved by testing the internal electric field with orthogonal tangent solenoidal disks. We demonstrate that a Method-of-Moments (MoM) discretization of the Curl-EFIE converges to a Galerkin-discretized MFIE as the testing disk dimensions vanish, yielding stable, breakdown-free impedance matrices for low frequencies and dense grids. Unlike the strongly singular kernels of the MFIE, the Curl-EFIE utilizes weakly singular kernels, significantly simplifying source integral evaluations. As a first-kind integral equation, it bypasses the MFIE's Gram matrix requirement, facilitating the analysis of non-matching triangulations. Furthermore, a linear combination of the Curl-EFIE and the conventional EFIE provides an interior-resonance-free formulation analogous to the Combined Field Integral Equation (CFIE). Finally, the Curl-EFIE performs particularly well at capturing the scattering behavior of sharp-edged and cornered geometries.**

## I. Introduction

THROUGH the Surface Equivalence Theorem, the scattering of an incident electromagnetic field by a perfectly conducting (PEC) object in free space can be modeled by equivalent boundary currents. The resulting scattered fields are then expressed as the spatial convolution of these currents with the free-space Green's function. The Electric-Field Integral Equation (EFIE) and Magnetic-Field Integral Equation (MFIE) [1] emerge from enforcing electric and magnetic boundary conditions, respectively, over the boundary surface enclosing the object. The equivalent currents are then obtained by solving either of these equations, which yield the same unique solution. However, at certain discrete frequencies, the homogeneous EFIE and MFIE admit non-trivial spurious resonant solutions. The Combined Field Integral Equation [2] (CFIE) is then constructed as a linear combination of the EFIE and MFIE, effectively yielding a strictly well-posed equation at any frequency.

The spectrum of the EFIE, a first-kind integral equation, diverges in two distinct directions. While the vector potential component is compact, leading to an eigenvalue accumulation point at zero, the scalar potential component is unbounded, with growing eigenvalues toward infinity. Upon discretization—conventionally via the Method of Moments (MoM) [3]— this spectral duality yields ill-conditioned impedance matrices —particularly at low frequencies or with fine meshes— that compromise the convergence of iterative solvers. Furthermore, the double-derivative nature of the scalar potential term yields a hypersingular Kernel that is normally mitigated numerically through a Galerkin MoM-discretization based on divergence-conforming sets [4] of basis functions (e.g. RWG [5]). A weakly singular form of the Kernel is then obtained, which allows the fast and accurate evaluation of the impedance elements [6−8].

The MFIE stands for a second-kind equation comprising an identity and a compact operator —strictly guaranteed for smooth manifolds [9]—, whereby the eigenvalues cluster tightly around a fixed point. The discretization of the MFIE thus gives rise to well-conditioned impedance matrices, allowing iterative solvers to reach convergence with significantly fewer iterations than the EFIE. However, the accurate matrix-element computation in the MFIE becomes rather elaborate, especially for high-order basis functions, because of the burdensome evaluation of the strongly quasi-singular Kernel contributions between near non-coplanar interactions [10−12]. Interestingly, The MFIE is compatible with a wide variety of basis function choices [13−15]. Due to its second-kind nature [16], the MFIE exhibits much better accuracy than the EFIE under high-order *p*-refinement for smooth, accurately modeled objects [17]. However, divergence-conforming MFIE MoM-schemes—though crucial for CFIE implementations—are typically less accurate than the EFIE when analyzing sharp-edged objects [13-15] or flat-faceted meshes over smooth surfaces [18].

In this paper, we present the new curl-based Electric-Field Integral Equation (Curl-EFIE) to evaluate electromagnetic scattering from PEC-bodies. Formulated by imposing a zero curl on the boundary EFIE—tested via orthogonally tangent solenoidal disks inside the body—this approach yields low-frequency and dense-grid stable MoM implementations that mimic a Galerkin-tested MFIE as the disk size approaches zero. Indeed, as we show in the paper, combining the Curl-EFIE and standard EFIE yields an interior-resonance-free, purely electric-field-based alternative to the traditional CFIE. Furthermore, the Curl-EFIE offers several distinct advantages over standard Galerkin MFIE MoM-schemes: (1) weakly singular kernels— in contrast to strongly singular kernels—which streamline impedance-element evaluations, especially for high-order basis functions or non-planar patches; (2) a first-kind integral equation formulation, as opposed to a second-kind equation, rendering it free of Gram-matrix computations; (3) enhanced RCS fidelity for sharp-edged geometries. Preliminary results have been presented in [19].

## II. Curl-based EFIE

We consider an incident electromagnetic wave $(\boldsymbol{E}_{\text{inc}}, \boldsymbol{H}_{\text{inc}})$ illuminating an impenetrable target, such as a PEC body. Let $\Omega_i \subset R^3$ denote the bounded volumetric domain defining the interior of a PEC body, and let $\Gamma = \partial\Omega_i$ represent its closed boundary surface. We denote the outward-pointing unit normal vector on the boundary as $\hat{\boldsymbol{n}}$. By virtue of Love's Equivalence Theorem, a solvable boundary-value formulation is achieved through domain homogenization (with free-space constants $\varepsilon_0$ and $\mu_0$), and the definition of a boundary-restricted equivalent electric surface current density $\boldsymbol{J}$. The tangential magnetic field $\boldsymbol{H}$ exhibits a jump discontinuity across the surface, directly proportional to the local surface current density ($\boldsymbol{J} = \hat{\boldsymbol{n}} \times \boldsymbol{H}$). The tangential component of the electric field $\boldsymbol{E}$ vanishes on $\Gamma$, maintaining continuity through a zero-value boundary condition into $\Omega_i$. The scattered fields $(\boldsymbol{E}_{\text{s}}, \boldsymbol{H}_{\text{s}})$ are defined as $(\boldsymbol{E} - \boldsymbol{E}_{\text{inc}}, \boldsymbol{H} - \boldsymbol{H}_{\text{inc}})$ in $\mathbb{R}^3 \setminus \Omega_i$ and as $(-\boldsymbol{E}_{\text{inc}}, -\boldsymbol{H}_{\text{inc}})$ in $\Omega_{\text{i}}$, because by the Equivalence Theorem, $(\boldsymbol{E}, \boldsymbol{H}) = \boldsymbol{0}$ throughout the interior volume.

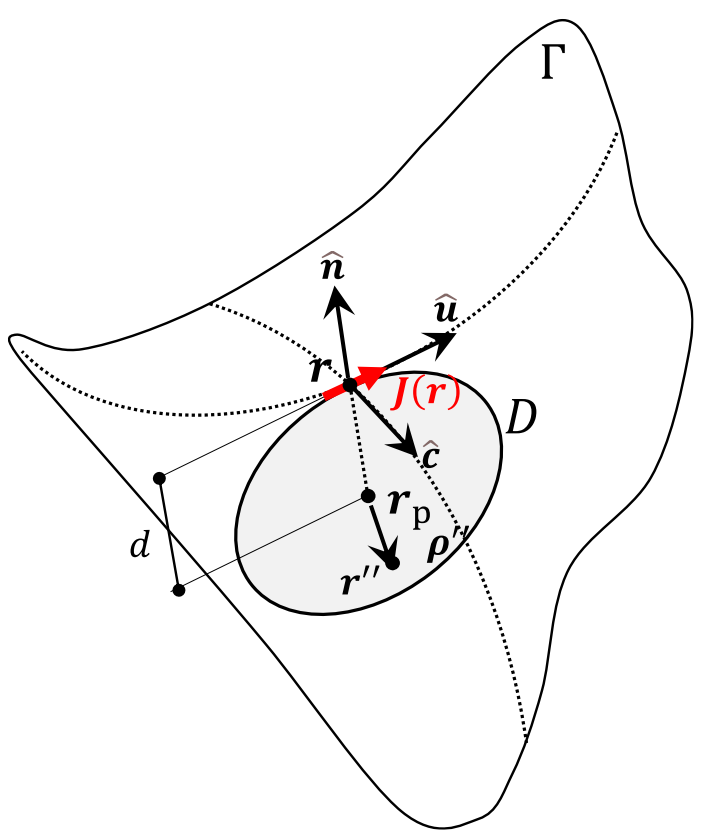

Fig. 1. Geometric location of the unit vectors $\{\hat{\boldsymbol{n}}, \hat{\boldsymbol{c}}, \hat{\boldsymbol{u}}\}$, and the testing disk $D$ with center $\boldsymbol{r}_{\text{p}}$ and radius $d$ associated with the point $\boldsymbol{r}$ at the boundary $\Gamma$ . $\hat{\boldsymbol{u}}$ is the unit vector parallel to the current at $\boldsymbol{r}, \boldsymbol{J}(\boldsymbol{r})$.

An interior point $\boldsymbol{r}_{\text{p}}(\boldsymbol{r})$, positioned in the vicinity of the boundary surface at a distance $d(\boldsymbol{r})$, centers a disk $D(\boldsymbol{r})$ tangent to the surface at $\boldsymbol{r}$ and with radius $d(\boldsymbol{r})$ (see Fig. 1). The disk is oriented to enclose the local surface current $\boldsymbol{J}(\boldsymbol{r})$, providing the local domain for enforcing the null-field condition in $\Omega_i$, $\boldsymbol{E}_s + \boldsymbol{E}_{\text{inc}} = 0$, via the surface integral

$$\iint_{D(\boldsymbol{r})} \overline{\boldsymbol{E}}_{\text{s}}(\boldsymbol{r}'') \cdot \boldsymbol{\psi}(\boldsymbol{r}, \boldsymbol{r}'')\, d\Gamma'' = -\iint_{D(\boldsymbol{r})} \overline{\boldsymbol{E}}_{\text{inc}}(\boldsymbol{r}'') \cdot \boldsymbol{\psi}(\boldsymbol{r}, \boldsymbol{r}'')\, d\Gamma'' \qquad \boldsymbol{r}'' \in D(\boldsymbol{r}), \boldsymbol{r} \in \Gamma \tag{1}$$

where $\boldsymbol{r}''$ stands for the coordinate vector spanning the disk surface $D(\boldsymbol{r})$ and $\overline{\boldsymbol{E}}_{\text{s}}$ ,$\overline{\boldsymbol{E}}_{\text{inc}}$ denote the ($jk$)-normalized scattered and incident fields: $\overline{\boldsymbol{E}}_{\text{s}} = \boldsymbol{E}_{\text{s}}/(-jk)$, $\overline{\boldsymbol{E}}_{\text{inc}} = \boldsymbol{E}_{\text{inc}}/(-jk)$ , where $k$ stands for the wavenumber and, under plane-wave incidence, $\overline{\boldsymbol{E}}_{\text{inc}} = j\hat{\boldsymbol{e}}_i e^{-jk\hat{\boldsymbol{k}}_i \cdot \boldsymbol{r}}/k$.

The testing function $\boldsymbol{\psi}(\boldsymbol{r}, \boldsymbol{r}'')$ is defined as a solenoidal vector field tangential to $D(\boldsymbol{r})$:

$$\boldsymbol{\psi}(\boldsymbol{r}, \boldsymbol{r}'') = \frac{1}{N_D} \hat{\boldsymbol{c}}(\boldsymbol{r}) \times \left(\boldsymbol{r}''(\boldsymbol{r}) - \boldsymbol{r}_{\text{p}}(\boldsymbol{r})\right) = \frac{1}{N_D}\left(\hat{\boldsymbol{c}}(\boldsymbol{r}) \times \boldsymbol{\rho}''(\boldsymbol{r})\right) \qquad \boldsymbol{r}'' \in D(\boldsymbol{r}), \boldsymbol{r} \in \Gamma \tag{2}$$

where $N_D = \pi d(\boldsymbol{r})^4/4$ and $\boldsymbol{\rho}''(\boldsymbol{r})$ denotes the local displacement vector from the center of the disk $\boldsymbol{r}_{\text{p}}(\boldsymbol{r})$. At contact point $\boldsymbol{r}$, the disk-normal unit vector $\hat{\boldsymbol{c}}(\boldsymbol{r})$ orients both $\boldsymbol{\psi}(\boldsymbol{r}, \boldsymbol{r}'')$ and the tangent unit vector $\hat{\boldsymbol{u}}(\boldsymbol{r})$ along the current direction $\boldsymbol{J}(\boldsymbol{r})$ (see Fig. 1). A right-hand basis set $\{\hat{\boldsymbol{n}}(\boldsymbol{r}), \hat{\boldsymbol{c}}(\boldsymbol{r}), \hat{\boldsymbol{u}}(\boldsymbol{r})\}$ is then established at $\boldsymbol{r}_{\text{p}}(\boldsymbol{r})$ such that the local displacement coordinates over the disk are $\boldsymbol{\rho}''(\boldsymbol{r}) = \left(n(\boldsymbol{r}), 0, u(\boldsymbol{r})\right)$. For the sake of brevity, the explicit dependence on the surface point $\boldsymbol{r}$ in the previous variables and functions is henceforth suppressed unless required for clarity (i.e. $(n, c, u) \equiv \left(n(\boldsymbol{r}), c(\boldsymbol{r}), u(\boldsymbol{r})\right), d \equiv d(\boldsymbol{r}), \boldsymbol{\psi}(\boldsymbol{r}'') \equiv \boldsymbol{\psi}(\boldsymbol{r}, \boldsymbol{r}''),\ \boldsymbol{\rho}'' \equiv \boldsymbol{\rho}''(\boldsymbol{r}), \hat{\boldsymbol{c}} \equiv \hat{\boldsymbol{c}}(\boldsymbol{r}), \hat{\boldsymbol{n}} \equiv \hat{\boldsymbol{n}}(\boldsymbol{r}), \hat{\boldsymbol{u}} \equiv \hat{\boldsymbol{u}}(\boldsymbol{r}), \boldsymbol{r}_{\text{p}} \equiv \boldsymbol{r}_{\text{p}}(\boldsymbol{r}), D \equiv D(\boldsymbol{r})$ ).

The expression in (1) evaluates the solenoidal first-order moments of the ($jk$)-normalized scattered and incident electric fields over $D$: $\mathcal{M}^D_{\boldsymbol{E}_{\text{s}}}(\boldsymbol{r})$, $\mathcal{M}^D_{\boldsymbol{E}_{\text{inc}}}(\boldsymbol{r})$. A linear-moment field condition is then obtained over the boundary of the scatterer

$$\mathcal{M}^D_{\boldsymbol{E}_{\text{s}}}(\boldsymbol{r}) + \mathcal{M}^D_{\boldsymbol{E}_{\text{inc}}}(\boldsymbol{r}) = \mathcal{M}^D_{\boldsymbol{E}}(\boldsymbol{r}) = \boldsymbol{0} \quad , \qquad \boldsymbol{r} \in \Gamma \tag{3}$$

where $\overline{\boldsymbol{E}} = \overline{\boldsymbol{E}}_{\text{s}} + \overline{\boldsymbol{E}}_{\text{inc}}$ denotes the total ($jk$)-normalized electric field.

A ($jk$)-normalized electric field $\overline{\boldsymbol{F}} = (F_n, F_c, F_u) \in \{\overline{\boldsymbol{E}}_s, \overline{\boldsymbol{E}}_{\text{inc}}\}$ at any coordinate $\boldsymbol{r}'' \in D$ ($c = 0$) may be represented by a multivariate Taylor expansion about $\boldsymbol{r}_{\text{p}}$ as

$$\overline{\boldsymbol{F}}(\boldsymbol{r}) = \overline{\boldsymbol{F}}(\boldsymbol{r}_{\text{p}}) + \left.\frac{\partial \overline{\boldsymbol{F}}}{\partial u}\right|_{\boldsymbol{r}_{\text{p}}} u + \left.\frac{\partial \overline{\boldsymbol{F}}}{\partial n}\right|_{\boldsymbol{r}_{\text{p}}} n + \boldsymbol{O}(|\boldsymbol{\rho}''|^2), \quad \boldsymbol{\rho}'' \in D \tag{4}$$

from which the linear contributions $\overline{\boldsymbol{F}}^{\text{ap}} \in \{\overline{\boldsymbol{E}}^{\text{ap}}_{\text{s}}, \overline{\boldsymbol{E}}^{\text{ap}}_{\text{inc}}\}$ are

$$\begin{aligned}\overline{\boldsymbol{F}}^{\text{ap}}(\boldsymbol{r}) = \overline{\boldsymbol{F}}(\boldsymbol{r}_{\text{p}}) &+ \left.\frac{\partial \bar{F}_c}{\partial n}\right|_{\boldsymbol{r}_{\text{p}}} n\hat{\boldsymbol{c}} + \left.\frac{\partial \bar{F}_c}{\partial u}\right|_{\boldsymbol{r}_{\text{p}}} u\hat{\boldsymbol{c}} + \left.\frac{\partial \bar{F}_n}{\partial n}\right|_{\boldsymbol{r}_{\text{p}}} n\hat{\boldsymbol{n}} \\ &+ \left.\frac{\partial \bar{F}_u}{\partial u}\right|_{\boldsymbol{r}_{\text{p}}} u\hat{\boldsymbol{u}} + \frac{1}{2}\,\mathcal{S}^D_{\boldsymbol{F}}(\boldsymbol{r}_{\text{p}}) \nabla''(nu) + \frac{1}{2}\mathfrak{D}^D_{\boldsymbol{F}}(\boldsymbol{r}_{\text{p}})(\hat{\boldsymbol{c}} \times \boldsymbol{\rho}'')\end{aligned} \tag{5}$$

where

$$\begin{aligned}\mathcal{S}^D_{\boldsymbol{F}}(\boldsymbol{r}) &\equiv \mathcal{S}^D_{\boldsymbol{F}}(\boldsymbol{r}_{\text{p}}) = \left[\frac{\partial \bar{F}_n}{\partial u} + \frac{\partial \bar{F}_u}{\partial n}\right]_{\boldsymbol{r}_{\text{p}}} \\ \mathfrak{D}^D_{\boldsymbol{F}}(\boldsymbol{r}) &\equiv \mathfrak{D}^D_{\boldsymbol{F}}(\boldsymbol{r}_{\text{p}}) = \left[\frac{\partial \bar{F}_n}{\partial u} - \frac{\partial \bar{F}_u}{\partial n}\right]_{\boldsymbol{r}_{\text{p}}}\end{aligned} \qquad \boldsymbol{r} \in \Gamma \tag{6}$$

and

$$\begin{aligned}\nabla''(nu) &= u\hat{\boldsymbol{n}} + n\hat{\boldsymbol{u}} \\ \hat{\boldsymbol{c}} \times \boldsymbol{\rho}'' &= u\hat{\boldsymbol{n}} - n\hat{\boldsymbol{u}}\end{aligned} \qquad \boldsymbol{\rho}'' \in D \tag{7}$$

Substituting (5) into (1) eliminates the first six terms

$$\bar{\boldsymbol{F}}(\boldsymbol{r}_{\mathrm{p}}) \cdot \hat{\boldsymbol{c}} \times \left( \iint_D \boldsymbol{\rho}''\, dS'' \right) = 0 \tag{8}$$

$$\begin{bmatrix} \partial \bar{F}_c / \partial u \\ \partial \bar{F}_c / \partial n \end{bmatrix}_{\boldsymbol{r}_{\mathrm{p}}} \iint_D \begin{bmatrix} u \\ n \end{bmatrix} \hat{\boldsymbol{c}} \cdot \boldsymbol{\psi}(\boldsymbol{r}'') dS'' = 0 \tag{9}$$

$$\begin{bmatrix} \partial \bar{F}_u / \partial u \\ \partial \bar{F}_n / \partial n \end{bmatrix}_{\boldsymbol{r}_{\mathrm{p}}} \iint_D \begin{bmatrix} u\hat{\boldsymbol{u}} \\ n\hat{\boldsymbol{n}} \end{bmatrix} \cdot \boldsymbol{\psi}(\boldsymbol{r}'')\, dS'' = \frac{1}{2N_D} \begin{bmatrix} -1 \\ +1 \end{bmatrix} \iint_D un\, dS'' = 0 \tag{10}$$

$$\frac{1}{2} \iint_D \nabla''(nu) \cdot \boldsymbol{\psi}(\boldsymbol{r}'')\, dS'' = \frac{1}{2N_D} \iint_D (u^2 - n^2)\, dS'' = 0 \tag{11}$$

reducing the expression to the final term

$$\frac{1}{2} \iint_D (\hat{\boldsymbol{c}} \times \boldsymbol{\rho}'') \cdot \boldsymbol{\psi}\, dS'' = \frac{1}{2N_D} \iint_D |\boldsymbol{\rho}''|^2\, dS'' = 1 \tag{12}$$

which yields

$$\left[ \mathfrak{D}^D_{\boldsymbol{E}_s}(\boldsymbol{r}) + \mathfrak{D}^D_{\boldsymbol{E}_{\mathrm{inc}}}(\boldsymbol{r}) \right] = \mathfrak{D}^D_{\boldsymbol{E}}(\boldsymbol{r}) = 0 \tag{13}$$

As $d \rightarrow 0$, such approximation recovers the conditions in (3) because the higher-order terms in (4) vanish relative to the stable linear contribution in (12)

$$\mathcal{M}^D_{\boldsymbol{E}}(\boldsymbol{r}) = \lim_{d \to 0} \mathfrak{D}^D_{\boldsymbol{E}}(\boldsymbol{r}) \quad , \quad \boldsymbol{r} \in \Gamma \tag{14}$$

## III. Magnetic-Field Correspondence

Using (6) we can transform (13) into:

$$\mathfrak{D}^D_{\boldsymbol{E}}(\boldsymbol{r}_{\mathrm{p}}) = \left[ \frac{\partial \bar{E}_n}{\partial u} - \frac{\partial \bar{E}_u}{\partial n} \right]_{\boldsymbol{r}_{\mathrm{p}}} = \hat{\boldsymbol{c}} \cdot [\nabla \times \bar{\boldsymbol{E}}]_{\boldsymbol{r}_{\mathrm{p}}} = 0 \tag{15}$$

which shows that the disk solenoidal testing in (1) extracts the $c$-component of the curl of the electric field at $\boldsymbol{r}_{\mathrm{p}}$.

From Faraday's law, $\nabla \times \bar{\boldsymbol{E}} = (\nabla \times \boldsymbol{E})/(-jk) = \eta_0 \boldsymbol{H}$, (15) follows as

$$\mathfrak{D}^D_{\boldsymbol{E}}(\boldsymbol{r}_{\mathrm{p}}) = \eta_0 \hat{\boldsymbol{c}} \cdot \boldsymbol{H}(\boldsymbol{r}_{\mathrm{p}}) = 0 \qquad \boldsymbol{r} \in \Gamma \tag{16}$$

or, equivalently,

$$\mathfrak{D}^D_{\boldsymbol{E}}(\boldsymbol{r}_{\mathrm{p}}) = \eta_0 \hat{\boldsymbol{c}} \cdot \left[ \boldsymbol{H}_{\mathrm{s}}(\boldsymbol{r}_{\mathrm{p}}) + \boldsymbol{H}_{\mathrm{inc}}(\boldsymbol{r}_{\mathrm{p}}) \right] = 0 \tag{17}$$

where $\eta_0$ denotes the free-space impedance. Therefore, the null interior magnetic-field condition in (17) relates to the curl-E condition in (1) via a multiplicative factor $\eta_0$. Consequently, the discretized electric-field operator $\mathfrak{D}^D_{\boldsymbol{E}}(\boldsymbol{r}_{\mathrm{p}})$ inherits the low-frequency and dense-mesh stability of MFIE implementations.

The electric-field expression in (1), which implements a disk testing touching the boundary, represents a first-kind integral and assumes a small non-zero value of $d$. On the other hand, the magnetic-field condition in (17) becomes second-kind in the limit $\boldsymbol{r}_p \rightarrow \boldsymbol{r}$ $(d \rightarrow 0)$. Specifically, the interior limit of the scattered magnetic $c$-field at the boundary, $\boldsymbol{H}_s(\boldsymbol{r})|_{\Gamma^-}$ comprises a jump term and a Cauchy-Principal-Value (CPV) integral, such that

$$\begin{aligned} \hat{\boldsymbol{c}} \cdot \boldsymbol{H}_{\mathrm{s}}(\boldsymbol{r})|_{\Gamma^-} &= -\frac{1}{2} \hat{\boldsymbol{c}} \cdot (\boldsymbol{J}(\boldsymbol{r}) \times \hat{\boldsymbol{n}}) + \hat{\boldsymbol{c}} \cdot \iint_{\Gamma, CPV} \boldsymbol{J}(\boldsymbol{r}') \times \nabla' G(\boldsymbol{r}, \boldsymbol{r}')\, d\Gamma' \\ &= -\frac{1}{2} \hat{\boldsymbol{u}} \cdot \boldsymbol{J}(\boldsymbol{r}) + \hat{\boldsymbol{c}} \cdot \iint_{\Gamma, CPV} \boldsymbol{J}(\boldsymbol{r}') \times \nabla' G(\boldsymbol{r}, \boldsymbol{r}')\, d\Gamma' \end{aligned} \tag{18}$$

which transforms the null magnetic field condition in (17) into the standard MFIE over the boundary surface $\Gamma$

$$\hat{\boldsymbol{c}} \cdot \left[ \iint_{\Gamma, CPV} \boldsymbol{J}(\boldsymbol{r}') \times \nabla' G(\boldsymbol{r}, \boldsymbol{r}')\, d\Gamma' + \boldsymbol{H}_{\mathrm{inc}}(\boldsymbol{r}_{\mathrm{p}}) \right] = \frac{1}{2} \hat{\boldsymbol{u}} \cdot \boldsymbol{J}(\boldsymbol{r}) \qquad \boldsymbol{r} \in \Gamma \tag{19}$$

## IV. Discretization of the Curl-EFIE

The expansion the electric current over the boundary $\boldsymbol{J}$ into an arbitrary set of basis functions $\{\boldsymbol{f}_n\}$ yields

$$\boldsymbol{J}(\boldsymbol{r}') \approx \tilde{\boldsymbol{J}}(\boldsymbol{r}') = \sum_{n=1}^{N} J(n)\, \boldsymbol{f}_n(\boldsymbol{r}') \quad , \quad \boldsymbol{r}' \in \Gamma_h \tag{20}$$

where $J(n)$ and $N$ are the current coefficients and total number of unknowns arising from the $h$-parameter meshed surface $\Gamma_h$. While the EFIE typically requires divergence-conforming basis functions, Curl-EFIE MoM-schemes —much like the MFIE— place no restrictions on the basis function choice. In this paper, we present Curl-EFIE results for several basis functions: edge-based RWG [5] (edge-normal continuous); facet-based monopolar-RWG [14] (edge-discontinuous) (see Fig. 2); and first-order Nodal Lagrangian Vector bases [20] (edge-discontinuous) (see Fig. 3).

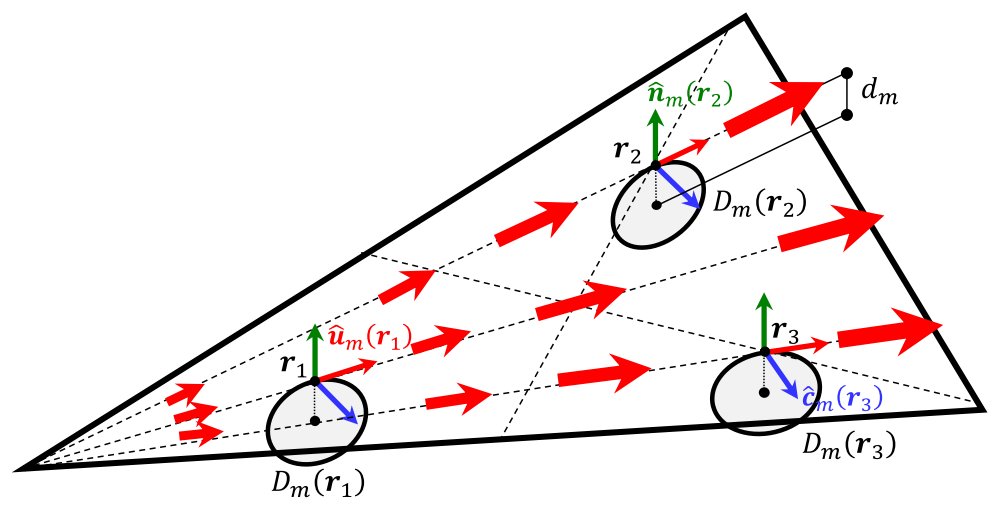

Fig. 2. Orientation of testing disks for a 3-point $\{\boldsymbol{r}_1, \boldsymbol{r}_2, \boldsymbol{r}_3\}$ quadrature rule associated with the $m$-th monopolar-RWG basis functions.

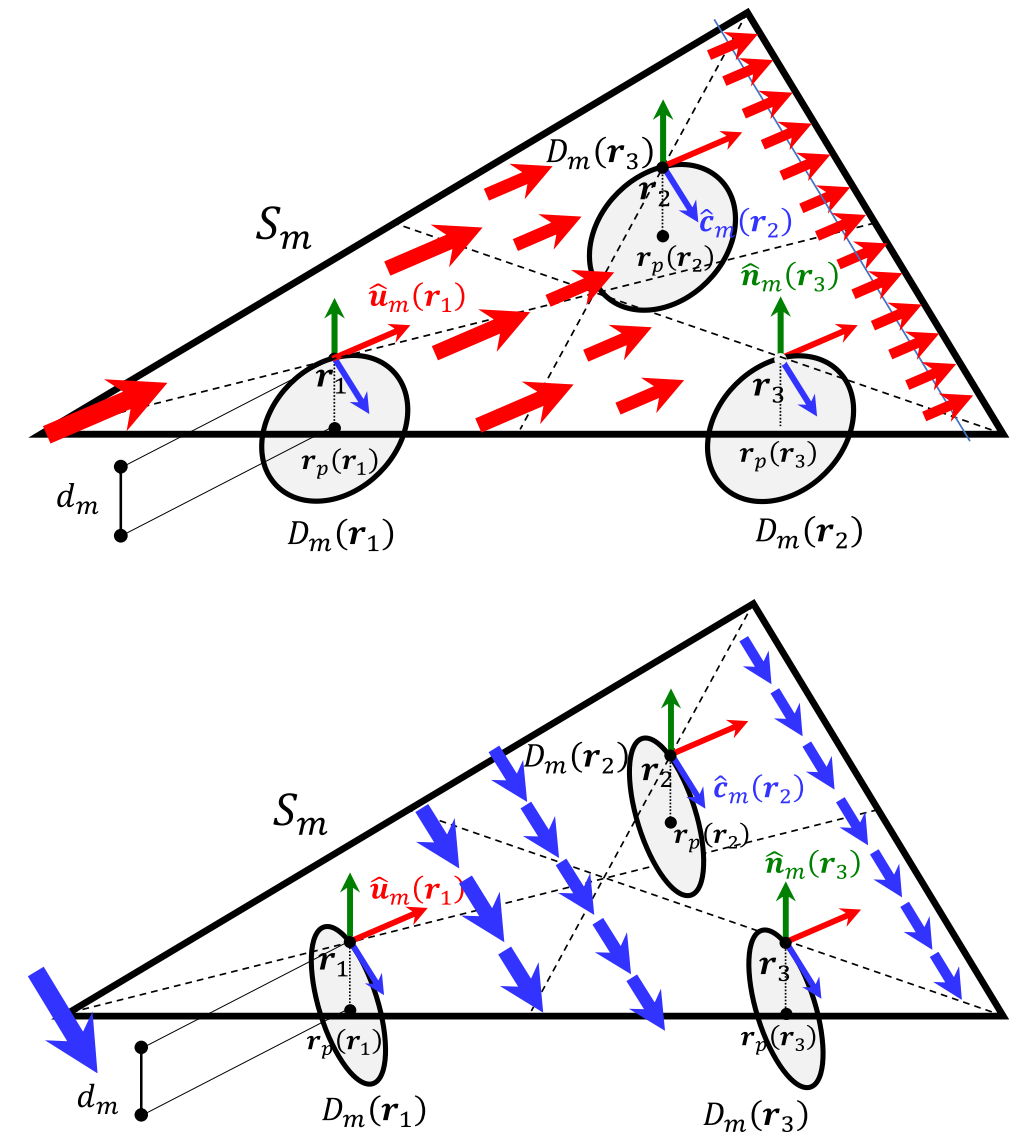

Fig. 3. Orientation of testing disks for a 3-point $\{\boldsymbol{r}_1, \boldsymbol{r}_2, \boldsymbol{r}_3\}$ quadrature rule associated with the $m$-th first-order nodal lagrangian basis function, oriented [parallel / perpendicular] to the edge opposite its associated node.

Our MoM-implementation of the Curl-EFIE is physically motivated by the alignment of the field-testing disks with the surface current and follows the Galerkin preference in standard EFIE MoM-schemes. We define the approximation of the first-order ($jk$)-normalized electric moment over the disk linked to the $m$-th basis function tangent to the point $\boldsymbol{r}$, $D_m \equiv D_m(\boldsymbol{r})$, as

$$\widetilde{\mathcal{M}}_{\boldsymbol{E}_s}^{D_m}(\boldsymbol{r}) = \iint_{D_m} \widetilde{\widetilde{\boldsymbol{E}}}_s(\boldsymbol{r}'') \cdot \boldsymbol{\psi}_m(\boldsymbol{r},\boldsymbol{r}'')\, d\Gamma'' \qquad \boldsymbol{r}'' \in D_m \;, \boldsymbol{r} \in \Gamma_h \tag{21}$$

The solenoidal testing function $\boldsymbol{\psi}_m$ over $D_m$ is defined as

$$\begin{aligned}\boldsymbol{\psi}_m(\boldsymbol{r},\boldsymbol{r}'') &= \frac{1}{N_D^m}\hat{\boldsymbol{c}}_m(\boldsymbol{r}) \times \left(\boldsymbol{r}'' - \boldsymbol{r}_p^m(\boldsymbol{r})\right) \\ &= \frac{1}{N_D^m}\hat{\boldsymbol{c}}_m \times \boldsymbol{\rho}''_m \qquad \boldsymbol{r}'' \in D_m \;, \boldsymbol{r} \in \Gamma_h\end{aligned} \tag{22}$$

where $\boldsymbol{r}_p^m \equiv \boldsymbol{r}_p^m(\boldsymbol{r})$ and $\hat{\boldsymbol{c}}_m \equiv \hat{\boldsymbol{c}}_m(\boldsymbol{r})$ represent, respectively, the geometric center of $D_m$ and the unit-vector oriented normally to the disk such that $\hat{\boldsymbol{c}}_m = \hat{\boldsymbol{u}}_m \times \hat{\boldsymbol{n}}_m$. The unit-vector $\hat{\boldsymbol{u}}_m \equiv \hat{\boldsymbol{u}}_m(\boldsymbol{r})$ is aligned with the testing function $\boldsymbol{f}_m$ at $\boldsymbol{r}$ at the meshed boundary (i.e. $\boldsymbol{f}_m(\boldsymbol{r}) = |\boldsymbol{f}_m(\boldsymbol{r})|\hat{\boldsymbol{u}}_m$ ) and $\hat{\boldsymbol{n}}_m$ stands for the outward-pointing unit normal vector at the corresponding facet (see Figs. 2 and 3). The radius of $D_m$, $d_m \equiv d_m(\boldsymbol{r})$, sets the normalizing factor $N_D^m$ to $\pi {d_m}^4/4$. The approximated scattered electric field $\widetilde{\widetilde{\boldsymbol{E}}}_s(\boldsymbol{r}'')$ in (21) results from the spatial convolution of the free-space Green's function $G$ with the approximated current $\tilde{\boldsymbol{J}}$ so that

$$\begin{aligned}\widetilde{\widetilde{\boldsymbol{E}}}_s(\boldsymbol{r}'') &= \eta_0 \sum_{n=1}^{N} J_n \iint_{S_n} G(\boldsymbol{r}'',\boldsymbol{r}')\boldsymbol{f}_n(\boldsymbol{r}')\, d\Gamma' \\ &+ \frac{1}{jk}\sum_{n=1}^{N} J_n \nabla''\widetilde{\Phi}_n(\boldsymbol{r}'') \qquad \boldsymbol{r}'' \in D_m\end{aligned} \tag{23}$$

where $G(\boldsymbol{r}'',\boldsymbol{r}') = e^{-jk|\boldsymbol{r}''-\boldsymbol{r}'|}/(4\pi|\boldsymbol{r}'' - \boldsymbol{r}'|)$ and $\nabla\widetilde{\Phi}_n$ denotes the contribution of the $n$-th basis function to the approximated potential gradient.

The evaluation of (21) with (23) yields

$$\widetilde{\mathcal{M}}_{\boldsymbol{E}_s}^{D_m}(\boldsymbol{r}) = \eta_0 \sum_{n=1}^{N} J_n \iint_{D_m} \boldsymbol{\psi}_m(\boldsymbol{r},\boldsymbol{r}'') \cdot \left[\iint_{S_n} G(\boldsymbol{r}'',\boldsymbol{r}')\boldsymbol{f}_n(\boldsymbol{r}')\, d\Gamma'\right] d\Gamma'' \quad , \; \boldsymbol{r} \in \Gamma_h \tag{24}$$

where it is satisfied [19]

$$\iint_{D_m} \boldsymbol{\psi}_m(\boldsymbol{r},\boldsymbol{r}'') \cdot \nabla\widetilde{\Phi}_n(\boldsymbol{r})dS'' = 0 \tag{25}$$

because the swirling testing vectors $\boldsymbol{\rho}''_m \times \hat{\boldsymbol{c}}_m$ have zero divergence and no leakage out of $D_m$ .

We construct the MoM-discretization of the Curl-EFIE formulation by testing the approximated linear-moment quantities in (21) over $\Gamma_h$ so that

$$\iint_{S_m} |\boldsymbol{f}_m(\boldsymbol{r})|\widetilde{\mathcal{M}}_{\boldsymbol{E}_s}^{D_m}(\boldsymbol{r})d\Gamma = -\iint_{S_m} |\boldsymbol{f}_m(\boldsymbol{r})|\mathcal{M}_{\boldsymbol{E}_{\text{inc}}}^{D_m}(\boldsymbol{r})d\Gamma \qquad 1 \le m \le N \tag{26}$$

which gives rise to the following system of equations

$$\sum_{n=1}^{N} Z^C(m,n)J(n) = -E_i^C(m) \;, \quad 1 \le m \le N \tag{27}$$

where the impedance-matrix and excitation vector elements, $Z^C(m,n)$ and $E_i(m)$, are defined as

$$Z^C(m,n) = \eta_0 \iint_{S_m} |\boldsymbol{f}_m(\boldsymbol{r})| \iint_{D_m} \boldsymbol{\psi}_m(\boldsymbol{r},\boldsymbol{r}'') \cdot \left[\iint_{S_n} G(\boldsymbol{r}'',\boldsymbol{r}')\boldsymbol{f}_n(\boldsymbol{r}')\, d\Gamma'\right] d\Gamma'' d\Gamma \tag{28}$$

$$E_i^C(m) = \iint_{S_m} |\boldsymbol{f}_m(\boldsymbol{r})| \iint_{D_m} \boldsymbol{\psi}_m(\boldsymbol{r},\boldsymbol{r}'') \cdot \overline{\boldsymbol{E}}_{\text{inc}}(\boldsymbol{r}'')d\Gamma'' d\Gamma \tag{29}$$

and $S_m$, $S_n$ denote the domains of $\boldsymbol{f}_m$ and $\boldsymbol{f}_n$ respectively.

Conversely, the standard Galerkin discretization of the MFIE arises from assigning (19) to the testing components $\hat{\boldsymbol{u}}_m$ and $\hat{\boldsymbol{c}}_m$ in the discretized space, or equivalently $\boldsymbol{f}_m$ and $\boldsymbol{f}_m \times \hat{\boldsymbol{n}}_m$, yielding the matrix equation

$$\sum_{n=1}^{N} Z^H(m,n)J(n) = -H_i(m) \;, \quad 1 \le m \le N \tag{30}$$

where

$$\begin{aligned}Z^H(m,n) = &\iint_{S_m} (\boldsymbol{f}_m(\boldsymbol{r}) \times \hat{\boldsymbol{n}}_m) \cdot \iint_{S_n,CPV} \boldsymbol{f}_n(\boldsymbol{r}') \times \nabla' G(\boldsymbol{r},\boldsymbol{r}')\, d\Gamma'\, d\Gamma \\ &- \frac{1}{2}\iint_{S_m} \boldsymbol{f}_m(\boldsymbol{r}) \cdot \boldsymbol{f}_n(\boldsymbol{r})\, d\Gamma\end{aligned} \tag{31}$$

$$H_i(m) = \iint_{S_m} (\boldsymbol{f}_m(\boldsymbol{r}) \times \hat{\boldsymbol{n}}_m) \cdot \boldsymbol{H}_{\text{inc}}(\boldsymbol{r})d\Gamma \tag{32}$$

and $G(m,n) = \iint_{S_m} \boldsymbol{f}_m(\boldsymbol{r}) \cdot \boldsymbol{f}_n(\boldsymbol{r})\, d\Gamma = \langle \boldsymbol{f}_p, \boldsymbol{f}_q \rangle$ stands for a Gram-matrix element given by the $L^2$ – inner product, a form resulting from the second-kind nature of the MFIE.

As demonstrated in section III, the impedance elements in (28) converge to the Galerkin MoM-entries of the MFIE in (31) with vanishing testing disk dimensions. In the MFIE, the mutual interaction between two basis functions with shared support decomposes into a Cauchy Principal Value integral and an impulsive residue—representing the local field jump at coincidence ($\boldsymbol{r} = \boldsymbol{r}'$)—which yields the Gram-matrix element in (31). Conversely, the curl-EFIE in (28) contains a weakly singular kernel ($R^{-1}$) that is integrable at source-field coincidence. On the other hand, evaluating the mutual interaction in (28) over a shrinking disk at $\boldsymbol{r}$—across the overlapping domain of the $p$-th and $q$-th basis functions—yields a dominant term $Z^\delta(p,q)$ defined by the so-called "residue contribution" $-\alpha(\boldsymbol{r})|(\boldsymbol{r} - \boldsymbol{r}'') \cdot \hat{\boldsymbol{n}}|$ [7]

$$Z^\delta(p,q) = \lim_{d_\cap \to 0} -\frac{\eta_0}{4\pi N_D^p} \iint_{S_\cap} \alpha(\boldsymbol{r})\left|\boldsymbol{f}_p(\boldsymbol{r})\right| \boldsymbol{f}_n(\boldsymbol{r}) \cdot \iint_{D_\cap} (\hat{\boldsymbol{c}}_p \times \boldsymbol{\rho}''_p)(d_\cap - n)d\Gamma'' d\Gamma \tag{33}$$

where $S_\cap$ stands for the shared support, $\alpha(\boldsymbol{r}) = 2\pi$ denotes the subtended solid angle at the facet, assumed flat, and $d_\cap$ stands for the radius of the testing disk $D_\cap$ at $\boldsymbol{r}$. Additionally, the evaluation of the disk integral in (33) produces

$$\iint_{D_\cap}(\hat{\boldsymbol{c}}_p\times\boldsymbol{\rho}_p'')(d_\cap-n)d\Gamma''=\hat{\boldsymbol{u}}_p\iint_{D_\cap}n^2d\Gamma''=N_D^p\hat{\boldsymbol{u}}_p \tag{34}$$

turning (33) into

$$\begin{aligned}Z^\delta(p,q)&=-\frac{\eta_0}{2}\iint_{S_\cap}\left|\boldsymbol{f}_p(\boldsymbol{r})\right|\boldsymbol{f}_q(\boldsymbol{r})\cdot\hat{\boldsymbol{u}}_p\ d\Gamma\\&=-\frac{\eta_0}{2}\iint_{S_\cap}\boldsymbol{f}_p(\boldsymbol{r})\cdot\boldsymbol{f}_q(\boldsymbol{r})\ d\Gamma=-\frac{\eta_0}{2}G(p,q)\end{aligned} \tag{35}$$

which corresponds with the $\eta_0$-normalized Gram-matrix term in (31).

While $Z^\delta(p,q)$ persists for finite testing disk dimensions, it becomes increasingly dominant over the mutual coplanar support as the disk $D_\cap$ shrinks [7]. Unlike the conventional MFIE, which requires exact surface testing to enforce the boundary condition, the Curl-EFIE facilitates a continuous transition from interior to surface behavior as the disk dimensions decrease. Consequently, this translates into a first-kind integral equation for the Curl-EFIE, contrasted with a second-kind formulation for the MFIE. Notably, the Curl-EFIE bypasses the explicit construction of the Gram matrix—a prerequisite for the MFIE—as its structure is implicitly inherent to the formulation. In practice, disk dimensions are chosen to be small relative to the mesh parameter yet bounded away from the vanishing limit. This yields an impedance matrix that, while slightly modified compared to the Galerkin-MFIE, offers superior numerical accuracy for sharp-edged geometries.

Mirroring the CFIE (which combines the EFIE and MFIE) [2], a linear combination of the Curl-EFIE and conventional EFIE leverages the Curl-EFIE–MFIE correspondence to yield an interior-resonance-free formulation. Since the CFIE's $\eta_0$-factor is already implicit in the Curl-EFIE—see (16) and (17)—our "curl-combined" MoM-implementation (Curl-CFIE) simplifies to the following system of equations in matrix form

$$\left(\mathbf{Z}^{\boldsymbol{E}}+\mathbf{Z}^{\mathrm{C}}\right)\ \mathbf{J}=-\left(\mathbf{E}_{\mathrm{i}}+\mathbf{E}_i^C\right) \tag{36}$$

where $\mathbf{Z}^{\mathrm{C}}=\left[Z^{\mathrm{C}}(m,n)\right]_{N\times N}$, $\mathbf{E}_i^C=\left[E_i^C(n)\right]_{N\times 1}$ and $\mathbf{J}=[J(n)]_{N\times 1}$. The impedance matrix $\mathbf{Z}^{\boldsymbol{E}}$ and the excitation vector $\mathbf{E}_{\mathrm{i}}$ arise from a Galerkin-discretization of the EFIE using the same basis functions as the Curl-EFIE. This choice typically requires divergence-conforming schemes, which are standard in EFIE MoM- schemes [5].

Unlike the dual-integral Galerkin MFIE in (31), our Curl-EFIE formulation in (28) relies on a triple surface integration. Typically, the multiple integrals in (28) and (31) are evaluated via a nested integration scheme: the outer integrals are computed numerically through standard Gaussian quadrature [21],[22], while the single, innermost source integral—where the kernel singularities reside—is handled explicitly via singularity subtraction [7][10][11] or singularity cancellation schemes [8][12]. According to (28), the vector-potential-based Curl-EFIE relies on weakly singular $O(R^{-1})$ source contributions, just like the divergence-conforming EFIE, which offers significant numerical advantages. In contrast, according to (31), the MFIE operates on the curl of the potential, introducing more demanding strongly singular $O(R^{-2})$ kernels. While $O(R^{-1})$ kernels allow singularity cancellation schemes based on normalized polar scalings, $O(R^{-2})$ kernels demand complex, nonlinear transformations. Similarly, singularity subtraction is more stable for $O(R^{-1})$ potentials because their simpler expansions and milder asymptotics prevent severe accuracy loss.

Driven by this nested structure, the practical Curl-EFIE implementation introduces a set of testing disks over the testing subdomain that match the quadrature points of the outermost integral. Since these disks are evaluated at each specific outer quadrature location, they are naturally oriented along the direction of the testing function at each point (see Figs.2 and 3) [19]. Furthermore, testing disks inside a particular facet share identical diameter $H$, scaled to a fraction of the average side length of the facet $h$. Numerical tests show that a minimal four-point quadrature rule for the disk surface integral in (28) provides very good accuracy. Therefore, in this work subsequent accuracy assessments between our Curl-EFIE and MFIE MoM-schemes utilize a variable number of points for the outermost field integrals in (28) and (31), paired with targeted singularity treatment for the innermost source integral while keeping the disk quadrature fixed at four points [22].

## V. Numerical results

### A. Accuracy

In Figs. 4−7, we assess the accuracy of the Curl-EFIE MoM-implementations using several basis functions: RWG (Figs. 4-7), monopolar-RWG (Figs. 4 and 5) and nodal linear Lagrangian basis functions (LIN) (Figs. 6 and 7). While the edge-based RWG set enforces normal continuity across edges, the monopolar-RWG and LIN sets are facet-based, discontinuous at edges. We compare the computed RCS for the Galerkin-discretized Curl-EFIE and MFIE using singularity subtraction for the impedance elements: near-dominant Kernel contributions in the innermost integrals are evaluated analytically, while the remaining source and outermost field integrals are computed via Gaussian quadrature. For simplicity, both numerical evaluations adopt the same number of sampling points (n). As an accuracy reference in Figs. 4–7, we use the RCS computed with a very fine RWG-EFIE discretization.

Monopolar-RWG functions (also known as half-RWG) match standard RWG functions at each facet but lack inter-element continuity constraints. This doubles the unknowns, resulting in three basis functions per facet (see Fig. 2). Furthermore, monopolar-RWG basis functions capture the sharp-edge effect in MFIE analysis better than the standard RWG set [14].

Using the triangle's barycentric coordinates $\xi_i$ (which equal one at their respective vertex and zero at the opposite edge), we define six LIN basis functions per facet as

$$\boldsymbol{s}^{u/l}{}_i(\boldsymbol{r})=\xi_i\begin{pmatrix}\hat{\boldsymbol{u}}_i\\\hat{\boldsymbol{l}}_i\end{pmatrix}\qquad i=1,2,3 \tag{37}$$

where $\hat{\boldsymbol{u}}_i$ and $\hat{\boldsymbol{l}}_i$ denote the normal and tangential unit vectors at the edge opposite the *i*-th vertex of the triangle (see Fig. 3). The Trintinalia-Ling (linear-linear) basis functions [23], which stand for the divergence-conforming counterparts to our discontinuous LIN functions, outperform the RWG-discretized MFIE for moderate or relatively large scattering problems [15],[24].

Figs. 4 and 5 show the average dB-deviation of the computed RCS against a very fine RWG-EFIE discretization for a 0.1λ-regular square pyramid and a 0.25λ-regular tetrahedron. These electrically small, sharp-edged targets rigorously evaluate MFIE accuracy under pronounced edge effects, which induce strong computationally demanding near-field kernel singularities. Results are plotted against the testing disk diameter ($H$) for the MFIE or Curl-EFIE MoM-schemes using RWG or monopolar-RWG basis functions and n=6 or n=9 quadrature points. Additionally, Figs. 4 and 5 include MFIE results using the Direct Evaluation Method (DEM) [25] for the extremely accurate computation of sharp-edge-adjacent mutual impedance elements. The DEM semi-analytically reduces the 4-D integrals in (31) to smooth 2-D integrals, allowing machine-precision in the impedance element evaluation. Thus, the DEM-computed MFIE RCS results serve as an accuracy reference that the singularity-subtracted MFIE (with n=6 or n=9 in Figs. 4 and 5) cannot achieve for the sharp-edged electrically small objects under test.

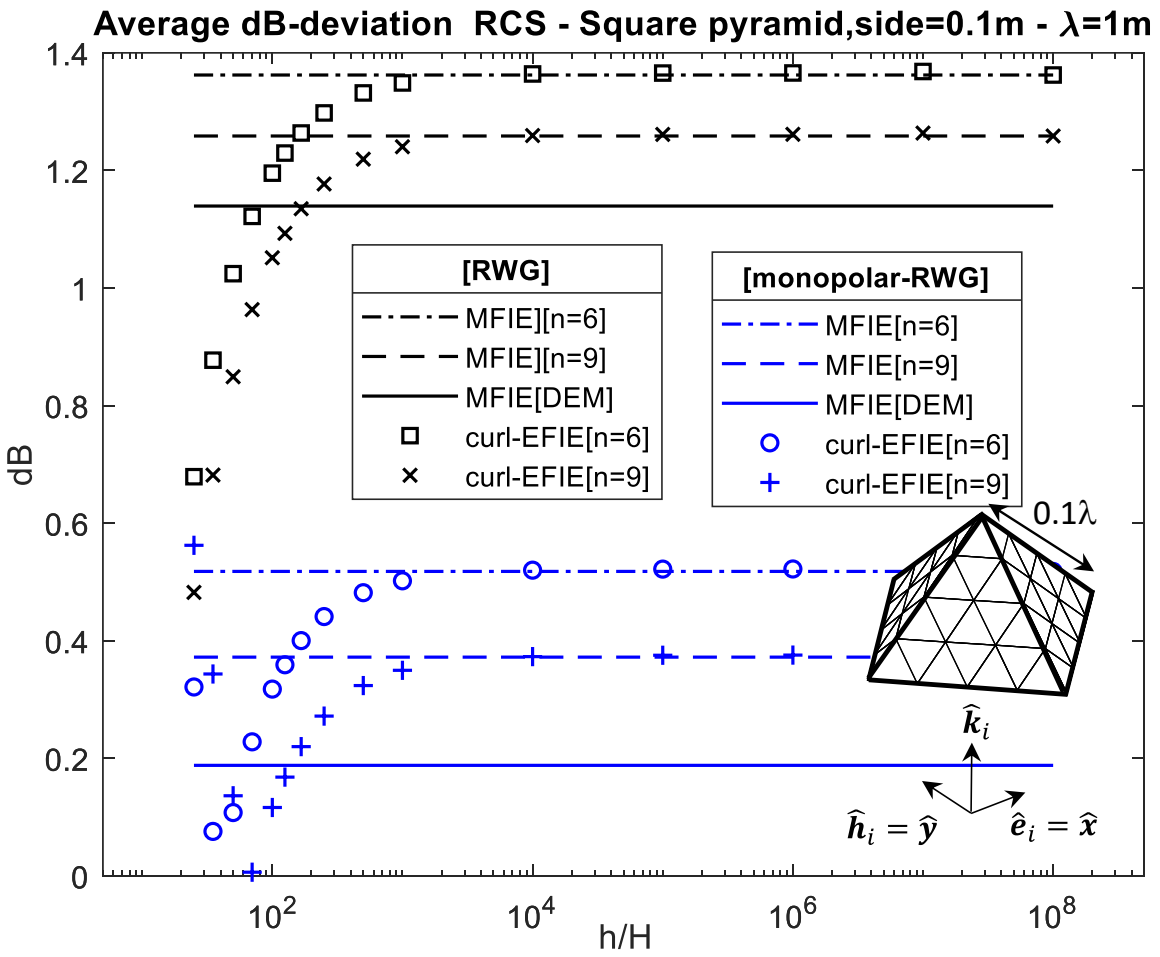

Fig. 4. Average RCS-deviation (300 MHz) over the E- and H-planes for a 0.1-m regular square pyramid meshed with 128 triangles under a z-propagating plane wave for several discretizations of the MFIE and the Curl-EFIE, with n=6 or n=9 quadrature points, versus the diameter $H$ of the testing disks, in terms of the mesh parameter $h$ . The reference is computed with the RWG-discretized EFIE using 66918 unknowns.

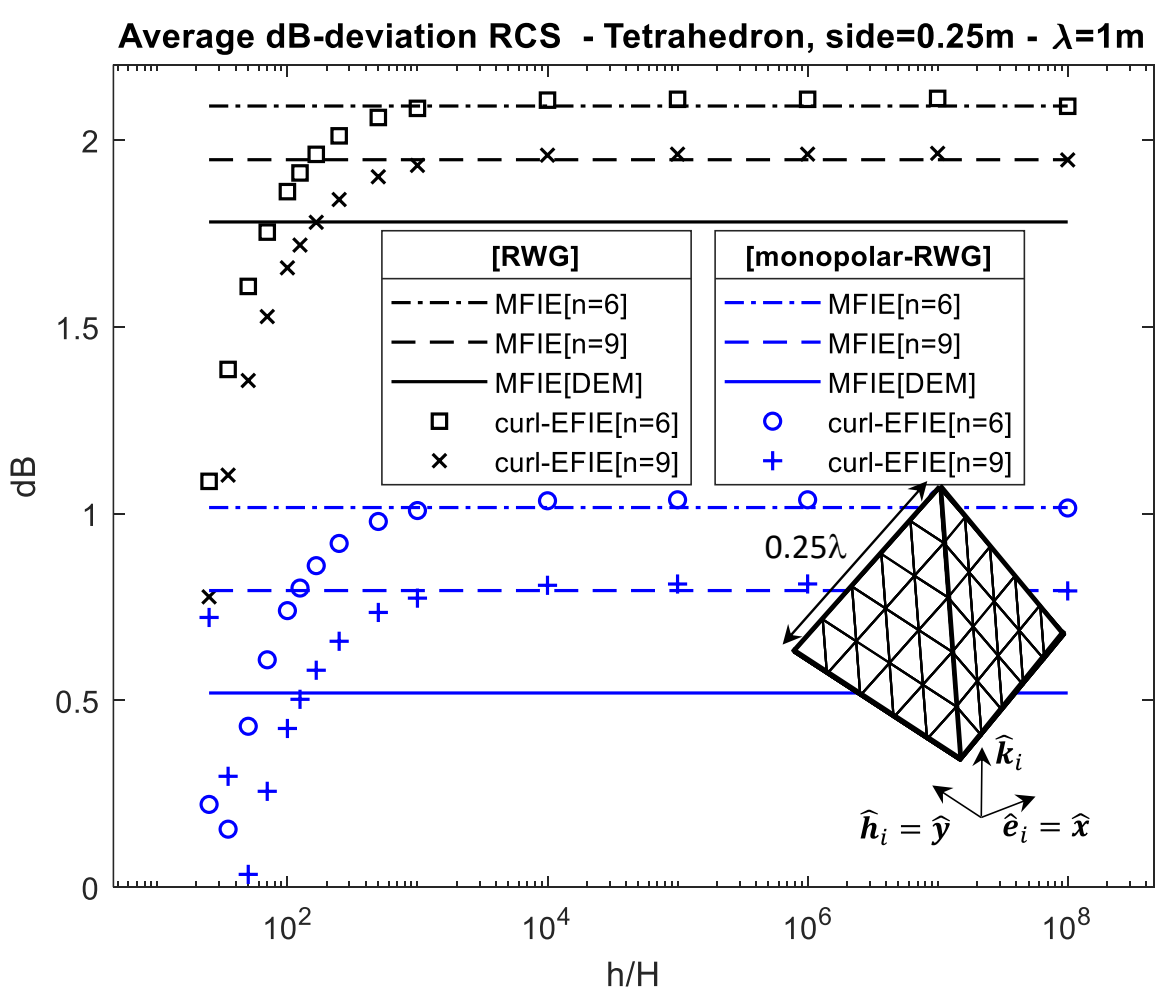

Fig. 5. Average RCS-deviation (300 MHz) over the E- and H-planes for a 0.25-m regular tetrahedron meshed with 100 triangles under a +z-propagating plane wave for several discretizations of the MFIE and the Curl-EFIE, with n=6 or n=9 quadrature points, versus the diameter $H$ of the testing disks, in terms of the mesh parameter $h$. The reference is computed with the RWG-discretized EFIE using 68475 unknowns.

As expected, the singularity-subtracted Curl-EFIE and MFIE schemes in Figs. 4 and 5 exhibit similar accuracy for $H \leq h/10^4$. Interestingly, testing-disk diameters around $h/10^2$ offer the best accuracy for both RWG and monopolar-RWG Curl-EFIE discretizations, even outperforming the MFIE under the DEM. This behavior aligns with the optimal $H$-values found when testing with prisms in the monopolar-RWG EFIE [26]. This further demonstrates that in-body, off-boundary testing—using disks or prisms attached to the surface— better captures edge-induced scattering in RCS calculation.

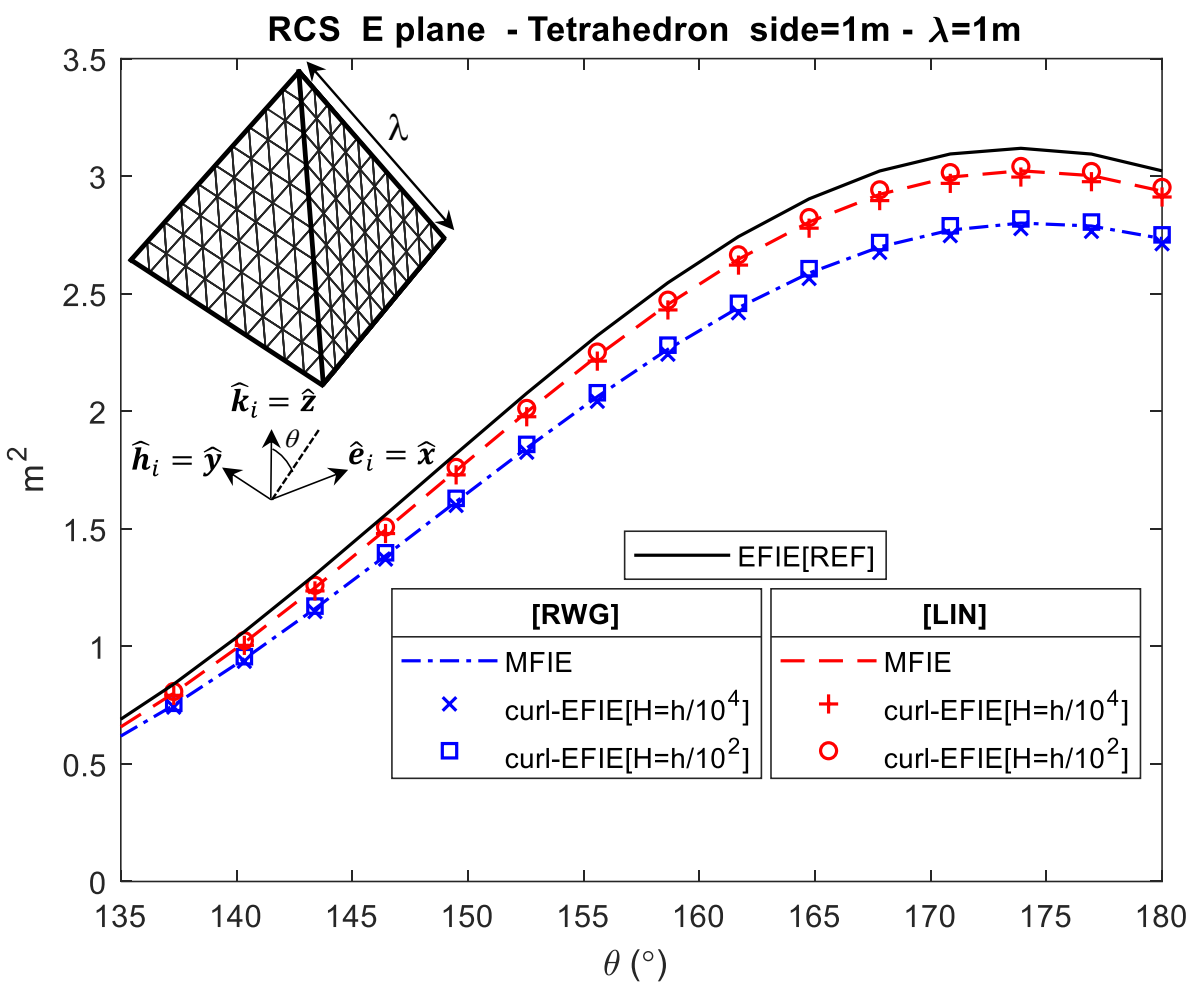

Fig. 6. E-plane bistatic RCS (300 MHz) for a 1-m regular tetrahedron meshed with 400 triangles under an x-polarized +z-propagating plane wave for RWG or LIN discretizations of the MFIE and the Curl-EFIE The reference is computed with the RWG-discretized EFIE using 71121 unknowns.

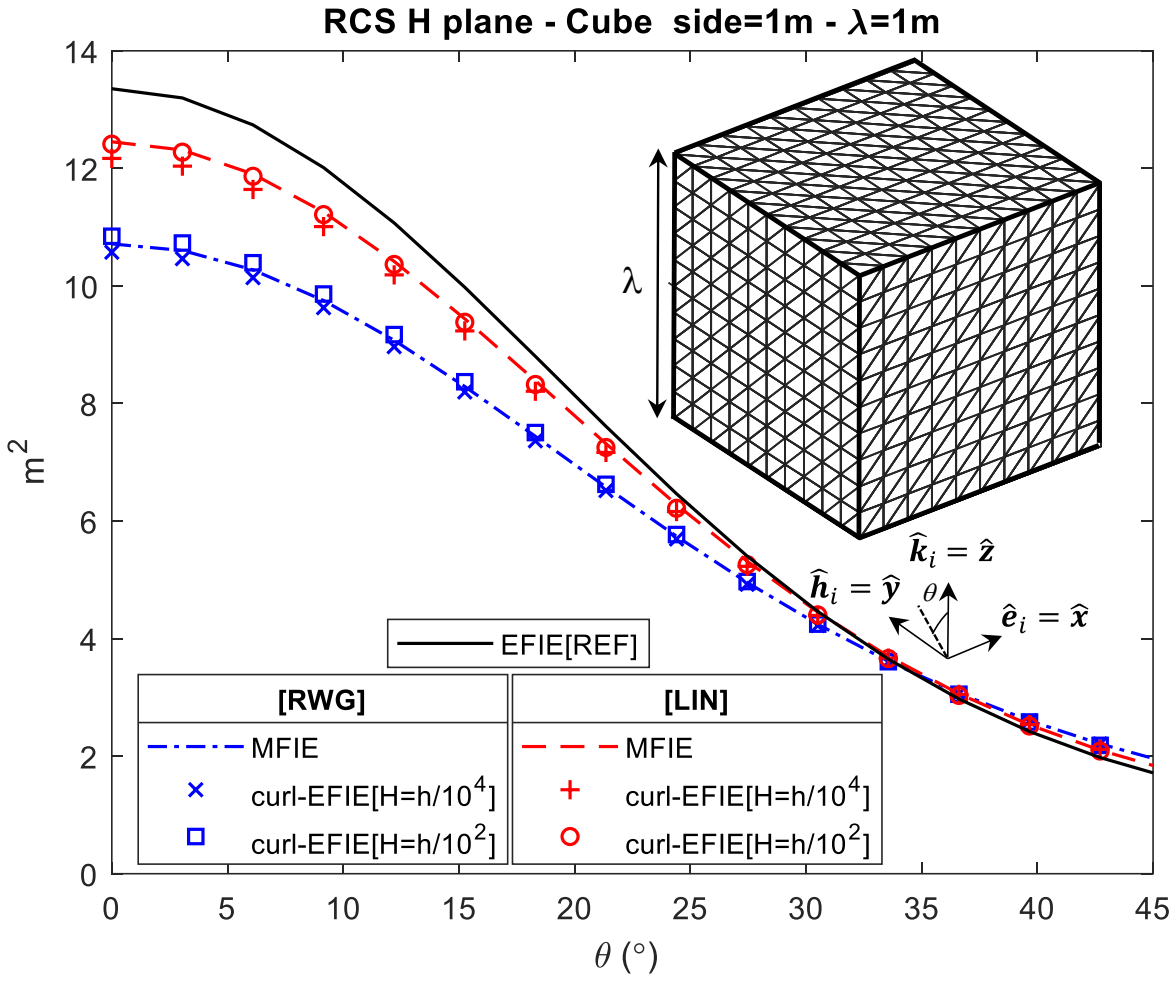

Fig. 7. H-plane bistatic RCS (300 MHz) for a 1-m cube meshed with 1200 triangles under an x-polarized +z-propagating plane wave incidence for RWG or LIN discretizations of the MFIE and the Curl-EFIE. The reference is computed with the RWG-discretized EFIE using 64701 unknowns.

Figs. 6 and 7 exhibit the computed RCS using singularity subtraction (n=6) for the MFIE and Curl-EFIE with RWG and LIN basis functions. The targets are electrically moderate, sharp-edged objects: a 1λ-regular tetrahedron and a 1λ-cube. Both Curl-EFIE schemes (RWG or LIN) show similar RCS performance across both $H = h/10^2$ or $H = h/10^4$ choices, though $H = h/10^2$ appears slightly better and subtly outperforms the MFIE, showing a minor average RCS improvement of 0.01 dB. This is likely because sharp edges influence the scattering of these 1λ-objects less than the subwavelength objects in Figs. 4 and 5.

While the singularity-subtracted RWG (or monopolar-RWG) MFIE schemes in Figs. 4 to 7 require analytical evaluation of the $1/R$, $(\boldsymbol{\rho}' - \boldsymbol{\rho})/R$, $1/R^3$, and $(\boldsymbol{\rho}' - \boldsymbol{\rho})/R^3$ integrals [7],[10],[11], the LIN-Galerkin variant (Figs. 6 and 7) further demands the more convoluted analytical evaluation of the $\xi_i/R^3$ Kernel term [11]. In contrast, the Curl-EFIE schemes, just like the EFIE, streamline this process, requiring only the $1/R$ and $(\boldsymbol{\rho}' - \boldsymbol{\rho})/R$ source Kernel contributions [7] in all tested cases.

## B. Non-matching meshes

Analyzing composite PEC structures composed of independently meshed, touching closed bodies is essential for versatile and flexible modular design. For such configurations, the first-kind electric field integral equation (EFIE) seamlessly handles cross-interactions between touching grids with different mesh densities (non-matching) as if the elements were perfectly aligned [27].

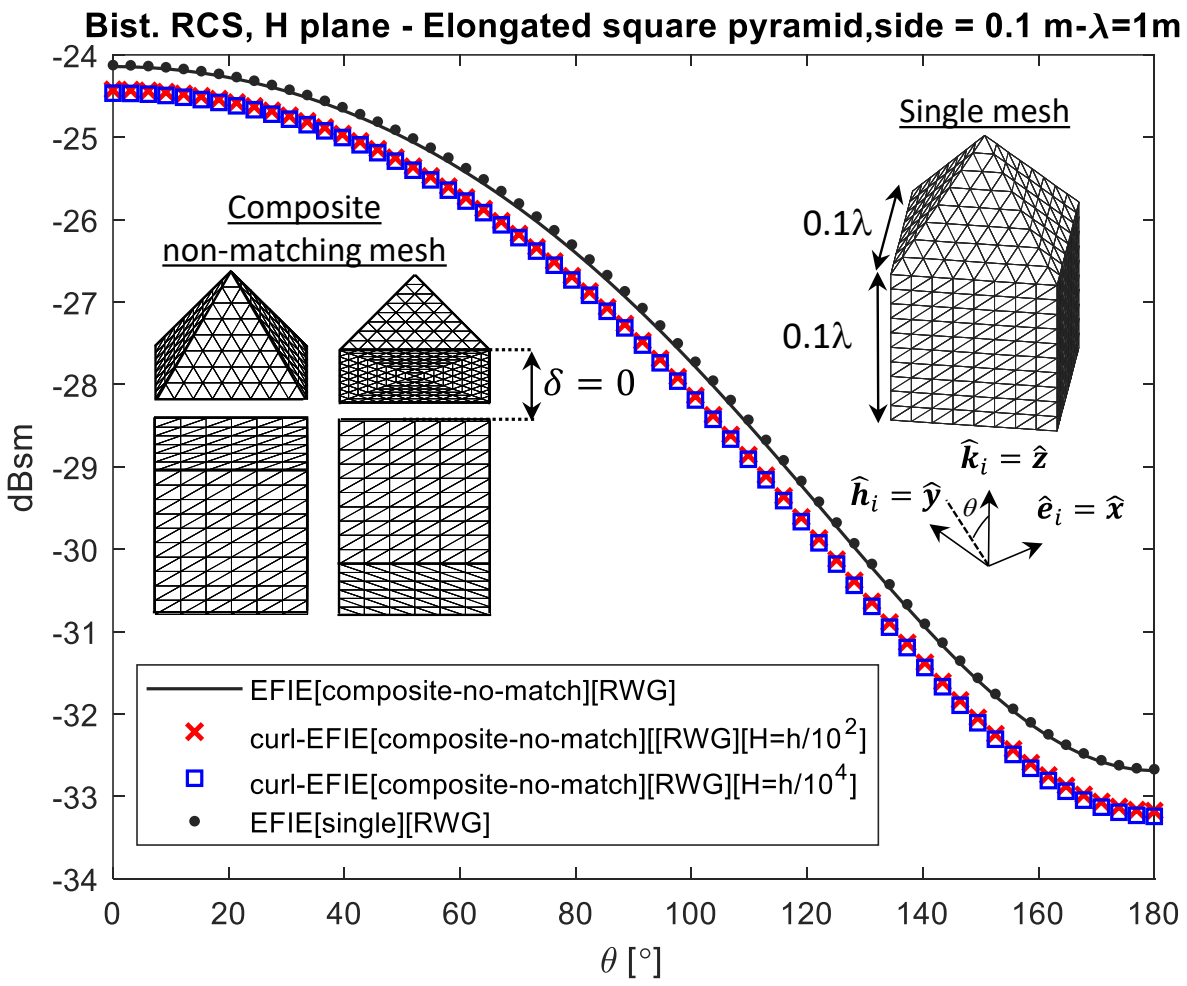

Fig. 8. H-plane bistatic RCS at 300 MHz for a 0.1-m elongated square pyramid under an x-polarized, +z-propagating plane wave, computed using the RWG-discretized EFIE and Curl-EFIE. The object is modeled either using a single mesh of 1152 triangles over the surface of the elongated square pyramid (single) or via the juxtaposition of non-matching meshes over a cube (624 triangles) and a square pyramid (512 triangles) for a total of 1136 triangles (composite-no-match).

Although the MFIE can theoretically handle non-matching triangulations, its practical implementation is significantly more complex than the EFIE and requires exceptional handling compared to standard interactions between non-touching facets. Indeed, Gram matrix calculation for mismatched touching meshes requires finding the overlapping regions between intersecting triangles and applying custom integration formulas [28],[29]. Furthermore, this MFIE evaluation demands precise alignment between the interfaces; even a minor numerical shift bypasses the Gram-matrix computation, triggering highly singular near-field interactions that are prone to inaccuracies.

Figure 8 displays the RCS computed via the RWG-discretized EFIE and Curl-EFIE for a 0.1λ-elongated square pyramid. The structure is modeled using two meshing approaches: a single mesh over the surface of the elongated square pyramid (single mesh), and two juxtaposed, independent, non-matching meshes for the cube and the square pyramid (composite non-matching mesh). Despite the mismatched interface in the latter case, the RWG-discretized Curl-EFIE yields RCS results that closely follow those obtained with the standard RWG-discretized EFIE on both cases. Again, Curl-EFIE with $H = h/10^2$ is slightly more accurate than with $H = h/10^4$, yielding a 0.05 dB average RCS improvement over the E and H planes. Like the EFIE, the first-kind Curl-EFIE applies identical treatment to both non-matching overlapping facets and non-touching interactions.

## C. Discussion

Based on Figs. 4-8, the Curl-EFIE outperforms the standard MFIE for near-field impedance matrix elements due a simplified treatment of both singular and quasi-singular integrals and the elimination of Gram-matrix contributions. Additionally, the Curl-EFIE appears better suited for analyzing sharp edges and corners, where the MFIE singular-Kernel treatment is particularly involved. Generally, intermediate- and far-field elements bypass singularity subtraction, requiring only standard quadrature. As the source-to-test distance increases, the number of quadrature points decreases until a single point suffices for the far-field impedance evaluation. While the Curl-EFIE in (28) involves a triple integral requiring a fixed four-point quadrature on the testing disks to evaluate $G$, the MFIE in (31) relies on a double integral featuring a cross product with $\nabla G$. Therefore, an unoptimized quadrature implementation for the Curl-EFIE in (28) incurs higher computational cost than the MFIE due to the fixed four-point testing integral, despite reducing the number of integration points in the remaining integrals. The Appendix presents an efficient approach for the Curl-EFIE quadrature, where a Taylor expansion at the disk center avoids evaluating the Green's function at each of the four testing points. This reduces the disk integral to a single central evaluation, making the computational cost of the Curl-EFIE for non-near impedance evaluations comparable to that of the MFIE. This Taylor-based quadrature may also be applied to low-order kernel contributions in near-field singularity subtractions, provided the source-to-test distance is much larger than $H$. This condition is strictly satisfied by our choice of $h/10^4 \leq H \leq h/10^2$.

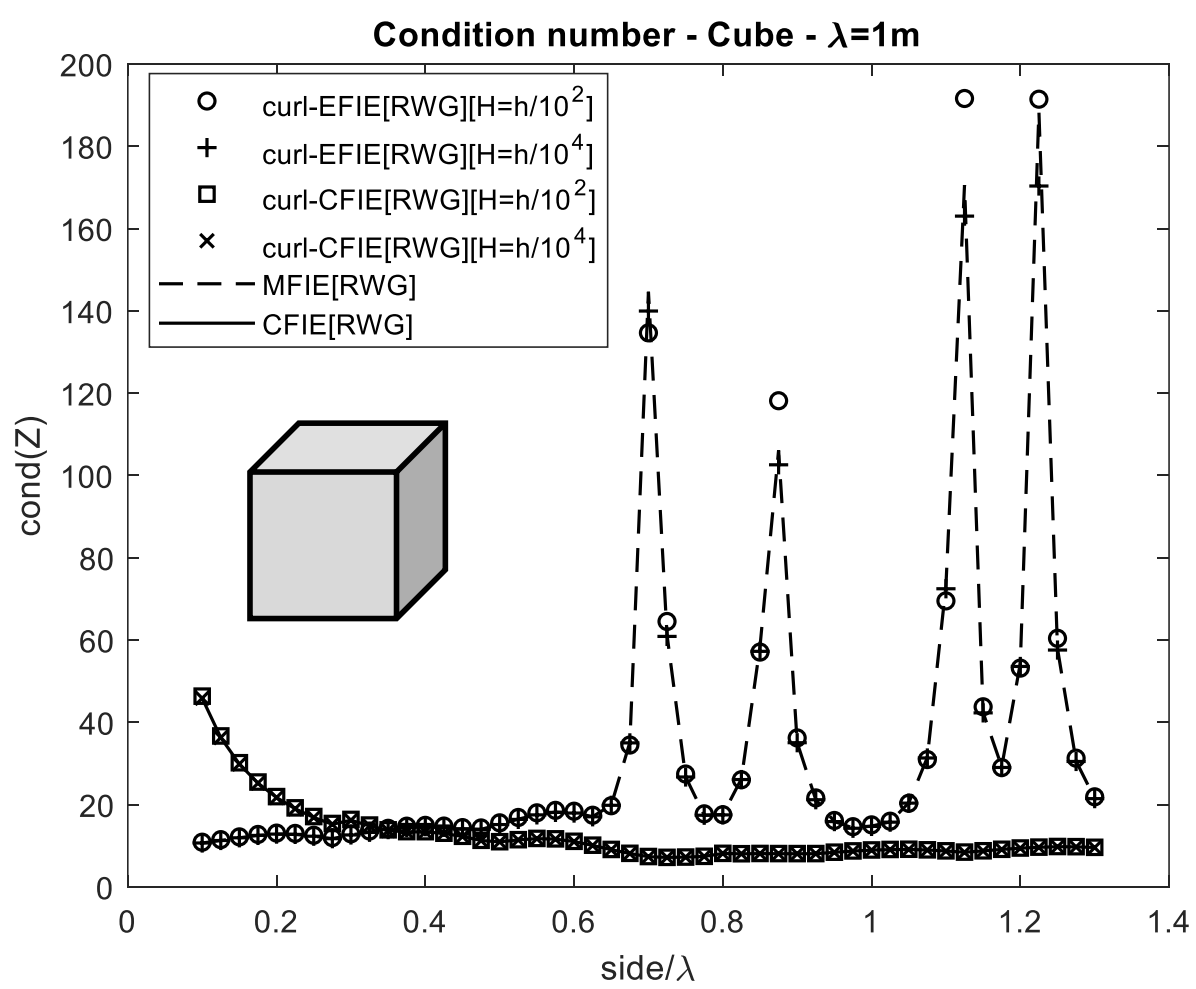

Fig. 9. Condition number of the impedance matrix resulting from the RWG-discretized Curl-EFIE, Curl-CFIE, MFIE and CFIE for a cube with growing electrical dimensions. The mesh size $h$ scales from 0.01λ to 0.09λ for side dimensions up to 0.9λ, stabilizing at approximately 0.1λ for larger dimensions.

### D. Curl-CFIE

Figure 9 demonstrates that the impedance matrix conditioning for the RWG-discretized Curl-CFIE (36) remains stable across frequencies, avoiding the numerical anomalies of the RWG-Galerkin Curl-EFIE and MFIE schemes. This confirms the Curl-CFIE as a resonance-free alternative to the standard CFIE. Figure 10 compares the RCS of a 2λ-cube computed via the RWG-discretized CFIE and Curl-CFIE (n=3) against the EFIE baseline. Curl-CFIE schemes exhibit similar RCS performance across both $H = h/10^2$ or $H = h/10^4$ choices, though the former yields slightly better results, very subtly outperforming the CFIE.

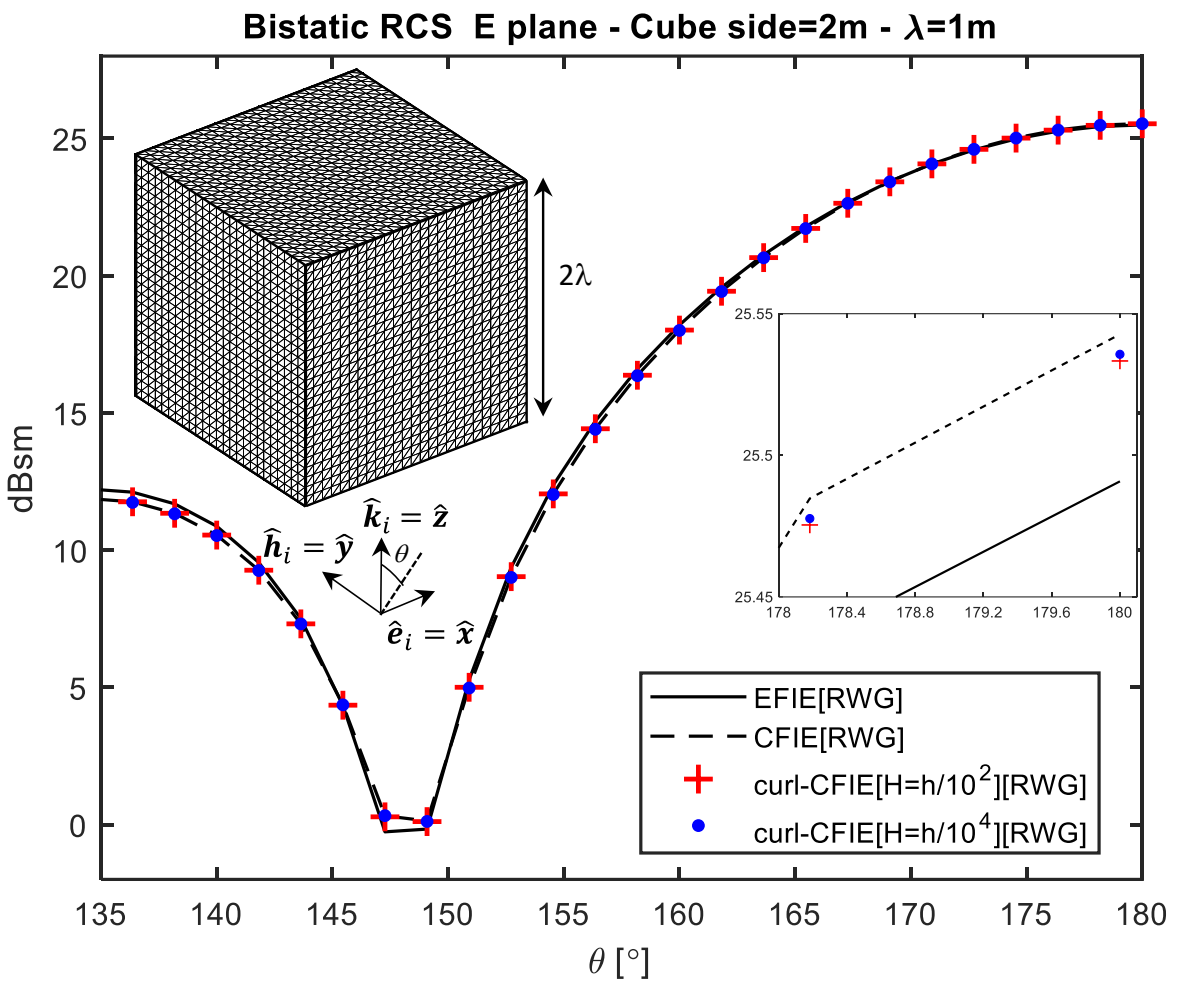

Fig. 10. E-plane bistatic RCS (300 MHz) for a 2-m cube meshed with 11250 triangles under an x-polarized +z-propagating plane wave incidence for RWG discretizations of the EFIE, the MFIE and the Curl-EFIE.

TABLE I
CONDITION NUMBER IN THE VERY LOW FREQUENCY REGIME OF THE IMPEDANCE MATRIX FOR A SPHERE WITH RADIUS = 1M MESHED WITH 128 TRIANGLES AND THE RWG-DISCRETIZED EFIE, MFIE AND CURL-EFIE

| | $\lambda$=10$^4$m | $\lambda$=10$^7$m | $\lambda$=10$^{10}$m | $\lambda$=10$^{13}$m |
|---|---|---|---|---|
| EFIE | $1.93 \times 10^8$ | $1.95 \times 10^{14}$ | $7.88 \times 10^{17}$ | $1.77 \times 10^{17}$ |
| MFIE | 3.513 | 3.513 | 3.513 | 3.513 |
| Curl-EFIE [$H$=$h$/10$^4$] | 3.514 | 3.514 | 3.514 | 3.514 |
| Curl-EFIE [$H$=$h$/10$^2$] | 3.591 | 3.591 | 3.591 | 3.591 |

TABLE II
CONDITION NUMBER FOR INCREASINGLY DENSE MESH DENSITIES OF THE IMPEDANCE MATRIX FOR A SPHERE WITH RADIUS = 0.1M AT 300 MHZ AND THE RWG-DISCRETIZED EFIE, MFIE AND CURL-EFIE

| | $h$=0.05λ | $h$=0.02λ | $h$=0.01λ | $h$=0.006λ |
|---|---|---|---|---|
| EFIE | 192.468 | 763.324 | $3.12 \times 10^3$ | $1.27 \times 10^4$ |
| MFIE | 3.769 | 3.784 | 3.799 | 3.847 |
| Curl-EFIE [$H$=$h$/10$^4$] | 3.766 | 3.783 | 3.799 | 3.847 |
| Curl-EFIE [$H$=$h$/10$^2$] | 3.742 | 3.755 | 3.794 | 3.846 |

### E. Stability

As demonstrated in Tables I and II for a 0.1 m sphere, the RWG-discretized Curl-EFIE maintains stable condition numbers for the $H \leq h/10^2$ range of interest at low frequencies and increasingly fine mesh densities, respectively. Due to its correspondence with the MFIE, the Curl-EFIE appears, as anticipated, immune to both low-frequency and dense-grid breakdowns. Furthermore, the condition number of the Curl-EFIE matrix using $H = h/10^4$ is closer to that of the MFIE than the choice $H = h/10^2$. This is expected because the Curl-EFIE impedance matrix asymptotically approaches the MFIE matrix as $H \to 0$ (see sections III and IV). Similarly, Tables I and II confirm that the EFIE condition number increases for decreasing frequencies and increasing meshing densities, respectively.

Figure 11 demonstrates the comparable RCS-accuracy of the RWG-discretized Curl-EFIE and MFIE schemes for electrically very small spheres (with a radius below $10^{-10}\lambda$), where the standard RWG-discretized EFIE fails due to the low-frequency breakdown [19],[30] (see red entries in Table I).

Additionally, the Curl-EFIE demands careful evaluation of the right-hand-side disk-testing integral in (27) under plane-wave incidence. In general, the zeroth-order terms of the incident field's complex exponential in (29) cancel completely when tested by the disk's solenoidal function. At the tiny electrical scales in Fig. 11, capturing this cancellation requires high-precision integration over the minuscule disk diameters. Since a naive quadrature evaluation suffers from catastrophic cancellation, we evaluate the disk integral in (29) as

$$\frac{j}{k}\iint_{D_m} (\boldsymbol{\psi}_m(\boldsymbol{r},\boldsymbol{r}'') \cdot \hat{\boldsymbol{e}}_i) \cdot \left(e^{-jk\hat{\boldsymbol{k}}_i \cdot \boldsymbol{r}} - 1\right) d\Gamma'' \tag{38}$$

which is readily possible in MATLAB via the built-in `expm1()` function.

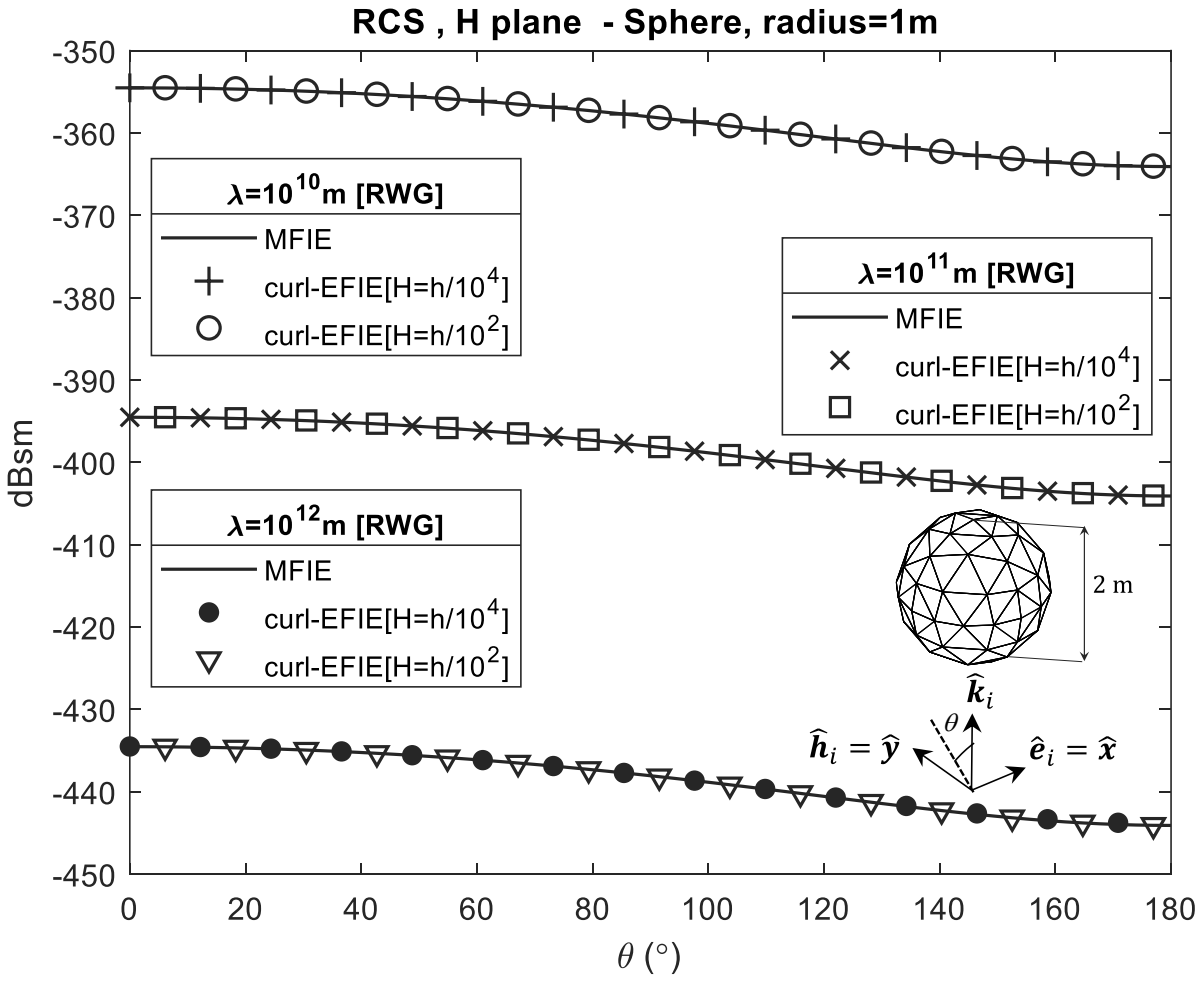

Fig. 11. H-plane bistatic RCS at $3 \times 10^{-2}$ Hz, $3 \times 10^{-3}$ Hz, $3 \times 10^{-4}$ Hz for a 1-m sphere meshed with 128 triangles under an x-polarized, +z-propagating plane wave, computed using the RWG-discretized MFIE and Curl-EFIE.

## VI. CONCLUSION

We present a curl-based electric-field integral equation (Curl-EFIE) designed for analyzing electromagnetic scattering from perfect electric conductors. This formulation is derived by enforcing a vanishing curl on the EFIE over the boundary manifold, which is implemented by testing the internal electric field with orthogonal tangent solenoidal disks. We show that when the testing disk dimensions approach zero, a Method-of-Moments (MoM) discretization of the Curl-EFIE converges to a Galerkin-discretized MFIE. This property yields stable, breakdown-free impedance matrices at low frequencies and on dense grids. Consequently, for sufficiently small testing-disk diameters (around or below $h/100$), the first-kind Curl-EFIE matches the spectral stability of a second-kind equation. Furthermore, as a first-kind integral equation, it avoids the Gram matrix requirement in MFIE implementations, simplifying non-matching triangulation analysis. Unlike the strongly singular kernels of the MFIE, the Curl-EFIE utilizes weakly singular kernels. This greatly simplifies source integral evaluations, benefiting impedance calculations for high-order basis functions or non-planar facets.

Our tests show that the Curl-EFIE with the disk-diameters around $h/100$ delivers enhanced RCS-accuracy, as compared with the MFIE, for geometries featuring sharp edges and corners. Lastly, linearly combining the Curl-EFIE with the conventional EFIE produces an interior-resonance-free formulation, the curl-CFIE, analogous to the Combined Field Integral Equation (CFIE).

## APPENDIX

### TAYLOR EXPANSION FOR THE CURL-EFIE QUADRATURE

The triple quadrature applied to the facet-to-facet contributions in $Z^{C}(m,n)$ yields

$$\eta_0 \sum_{i=1}^{Nf} \omega_i^{\mathrm{F}} |\boldsymbol{f}_m(\boldsymbol{F}_i)| \sum_{j=1}^{4} \omega_j^{D} \boldsymbol{\psi}(\boldsymbol{D}_{i,j}) \cdot \sum_{k=1}^{Ns} \omega_k^{S} G(\boldsymbol{D}_{i,j}, \boldsymbol{S}_k) \boldsymbol{f}_n(\boldsymbol{S}_k) \tag{39}$$

where the sets $\{\boldsymbol{F}_1, \cdots, \boldsymbol{F}_{Nf}\}$, $\{\boldsymbol{D}_1, \cdots, \boldsymbol{D}_4\}$, $\{\boldsymbol{S}_1, \cdots, \boldsymbol{S}_{Ns}\}$ *and* $\{\omega_1^F, \cdots, \omega_{Nf}^F\}$, $\{\omega_1^D, \cdots, \omega_4^D\}$, $\{\omega_1^S, \cdots, \omega_{Ns}^S\}$ stand for the quadrature points and area-weighted coefficients, respectively, on the field facet ($\in S_m$), the testing disks, and the source facet ($\in S_n$).

The Green's function at the $j$-th quadrature point of the disk $\boldsymbol{D}_{i,j}$—associated with the $i$-th quadrature point of the field triangle $\boldsymbol{F}_i$—can be expanded via a Taylor series around the disk center $\boldsymbol{P}_i$ as

$$G(\boldsymbol{D}_{i,j}, \boldsymbol{S}_k) \approx G(\boldsymbol{P}_i, \boldsymbol{S}_k) + \nabla G(\boldsymbol{P}_i, \boldsymbol{S}_k) \cdot (\boldsymbol{D}_{i,j} - \boldsymbol{P}_i) \tag{40}$$

The accuracy of this expansion relies on the condition $|\boldsymbol{D}_{i,j} - \boldsymbol{P}_i| \ll |\boldsymbol{P}_i - \boldsymbol{S}_k|$, which is systematically ensured since, by design, $H \leq 10^{-2}h$.

Inserting (40) into (39) cancels the first constant term of (40) because the disk-testing function $\boldsymbol{\psi}$ is solenoidal and yields

$$\eta_0 \sum_{i=1}^{Nf} \omega_i^{\mathrm{F}} |\boldsymbol{f}_m(\boldsymbol{F}_i)| \cdot \sum_{k=1}^{Ns} \omega_k^{S} \boldsymbol{f}_n(\boldsymbol{S}_k) \cdot \sum_{j=1}^{4} \omega_j^{D} [\nabla G(\boldsymbol{P}_i, \boldsymbol{S}_k) \cdot \Delta \boldsymbol{r}_j] \boldsymbol{\psi}(\boldsymbol{D}_j) \tag{41}$$

where $\Delta \boldsymbol{r}_j \equiv (\boldsymbol{D}_{\mathrm{i,j}} - \boldsymbol{P}_i)$ and $\boldsymbol{\psi}(\boldsymbol{D}_j) \equiv \boldsymbol{\psi}(\boldsymbol{D}_{i,j})$ because the testing disks on the flat facet have identical dimensions and same local solenoidal definitions.

Furthermore, by virtue of

$$[\nabla G(\boldsymbol{P}_i, \boldsymbol{S}_k) \cdot \Delta \boldsymbol{r}_j] \boldsymbol{\psi}(\boldsymbol{D}_j) = (\boldsymbol{\psi}(\boldsymbol{D}_j) \otimes \Delta \boldsymbol{r}_j) \nabla G(\boldsymbol{P}_i, \boldsymbol{S}_k) \tag{42}$$

(41) reduces to

$$\eta_0 \sum_{i=1}^{Nf} \omega_i^{\mathrm{F}} |\boldsymbol{f}_m(\boldsymbol{F}_i)| \sum_{k=1}^{Ns} \omega_k^{S} \boldsymbol{f}_n(\boldsymbol{S}_k) \cdot [\mathbf{M} \nabla G(\boldsymbol{P}_i, \boldsymbol{S}_k)] \tag{43}$$

where $\mathbf{M} \in \mathbb{R}^{3\times 3}$ is given by

$$\mathbf{M} = \sum_{j=1}^{4} \omega_j^{D} [\boldsymbol{\psi}(\boldsymbol{D}_j) \otimes \Delta \boldsymbol{r}_j] \tag{44}$$

which encapsulates the local geometry of all disks on the testing facet and can be precomputed and stored with the other facet-related geometric quantities.

The double quadrature in (43)—analogous to the MFIE double integral in (31)—simplifies because the disk projection space is two-dimensional, spanned by $\hat{\boldsymbol{u}}$ and $\hat{\boldsymbol{n}}$ (Figs. 2 and 3)

$$\mathbf{M} = [\hat{\boldsymbol{u}} \quad \hat{\boldsymbol{n}}] \mathbf{K} \begin{bmatrix} \hat{\boldsymbol{u}}^T \\ \hat{\boldsymbol{n}}^T \end{bmatrix} \tag{45}$$

where $\mathbf{K} \in \mathbb{R}^{2\times 2}$ is defined as

$$\mathbf{K} = \sum_{\boldsymbol{j}=\boldsymbol{1}}^{4} \omega_j^{D} \begin{bmatrix} \psi_u^j \\ \psi_n^j \end{bmatrix} [\Delta u_{\boldsymbol{j}} \quad \Delta n_{\boldsymbol{j}}] \tag{46}$$

owing to the local disk coordinate representation $\boldsymbol{\psi}(\boldsymbol{D}_{\mathrm{j}}) = (\psi_u^j, \psi_n^j)$ and $\Delta \boldsymbol{r}_j = (\Delta u_{\boldsymbol{j}}, \Delta n_{\boldsymbol{j}})$.

## ACKNOWLEDGMENT

This work was supported in part by MICIU/AEI/ 10.13039/501100011033/FEDER, UE, under Projects PID2023-146245OB-C21, Project PID2022-136869NB-C31, and PDC2022-133091-I00.

## REFERENCES

[1] A. F. Peterson, S. L. Ray, and R. Mittra, *Computational Methods for Electromagnetics*, ed. Wiley – IEEE press, New York, 1998.

[2] J. R. Mautz and R. F. Harrington, “H-field, E-field, and combined-field solutions for conducting bodies of revolution,” *Archiv für Elektronik und Übertragungstechnik* (Electronics and Communication), vol. 32, no. 4, 1978, pp. 157–164.

[3] Harrington, R. E., *Field Computation by Moment Method*, New York : Macmillan, 1968, pp. 82-89.

[4] R. D. Graglia, D. R. Wilton and A. F. Peterson, "Higher order interpolatory vector bases for computational electromagnetics", *IEEE Trans. Antennas Propag.*, vol. 45, no. 3, pp. 329-342, March 1997.

[5] S. M. Rao, D. R. Wilton and A. W. Glisson, "Electromagnetic scattering by surfaces of arbitrary shape", *IEEE Trans. Antennas Propag.*, vol. AP-30, no. 3, pp. 409-418, May 1982.

[6] M. G. Duffy, “Quadrature over a pyramid or cube of integrands with a singularity at a vertex,” *SIAM Journal on Numerical Analysis*, vol. 19, no. 6, pp. 1260-1262, Dec. 1982.

[7] D. R. Wilton, S. M. Rao, A. W. Glisson, D. H. Schaubert, O. M. Al-Bundak, and C. M. Butler, "Potential integrals for uniform and linear source distributions on polygonal and polyhedral domains," *IEEE Trans. Antennas Propag.*, vol. AP-32, no. 3, pp. 276-281, Mar. 1984.

[8] M. A. Khayat, and D. R. Wilton, “Numerical evaluation of singular and near-singular potential integrals,” *IEEE Trans. Antennas Propag.*, vol. 53, no. 10, pp. 3180-3190, Oct. 2005.

[9] G. C. Hsiao, and R. E. Kleinman, “Mathematical foundations for error estimation in numerical solutions of integral equations in

electromagnetics," *IEEE Trans. Antennas Propag.*, vol. 45, no. 3, pp. 316-328, Mar. 1997.
[10] R. E. Hodges and Y. Rahmat-Samii, "The evaluation of MFIE integrals with the use of vector triangle basis functions," *Microw. Opt. Technol. Lett.*, Vol.. 14, no. 1, pp. 9-14, Jan. 1997.
[11] R. D. Graglia, "On the numerical integration of the linear shape functions times the 3D's Green's function or its gradient on a plane triangle," *IEEE Trans. Antennas Propag.*, vol. 41, no. 10, pp. 1448-1455, Oct. 1993.
[12] F. Vipiana, D. R. Wilton, and W. A. Johnson," Advanced numerical schemes for the accurate evaluation of 4-D reaction integrals in the method of moments," *IEEE Trans. Antennas Propag.*, vol. 61, no. 11, pp. 5559-5566, nov. 2013.
[13] E. Ubeda, and J. M. Rius, "MFIE MoM-formulation with curl-conforming basis functions and accurate Kernel integration in the analysis of perfectly conducting sharp-edged objects," *Microw. Opt. Technol. Lett.*, vol. 44, no. 4, pp. 354-358, Feb. 2005.
[14] E. Ubeda and J. M. Rius, "Novel Monopolar MoM-MFIE discretization for the scattering analysis of small objects", *IEEE Trans. Antennas Propag.*, vol. 54, no. 1, pp. 50-57, Jan. 2006.
[15] Ö. Ergül, L. Gürel, "Linear-Linear basis functions for MLFMA solutions of magnetic-field and combined-field integral equations," *IEEE Trans. Antennas Propag.*, vol. 55, no. 4, pp. 1103-1110, Apr. 2007.
[16] K. E. Atkinson, "The numerical solution of integral equations of the second kind," ed. Cambridge University Press, Cambridge, UK, 2010.
[17] S. D. Gedney, "High-order method-of-moments solution of the scattering by three-dimensions PEC bodies using quadrature-based point matching," *Microw. Opt. Technol. Lett.*, vol. 29, no. 5, pp. 303-309, June 2001.
[18] J. Kornprobst , and T. F. Eibert, "Accuracy analysis of div-conforming hierarchical higher-order discretization schemes for the magnetic field integral equation," *IEEE J. Multiscale Multiphys. Comput. Tech.*, pp. 261-268, vol. 8, 2023.
[19] E. Ubeda, and J. M. Rius, "Low-frequency stable discretization of the electric field integral equation for arbitrary meshes," in Proc. IEEE Antennas Propag. Soc. Int. Symp., Detroit, MI, USA, Jul. 12–17, 2026.
[20] J. P. Berrut, and L. N. Trefethen, "Barycentric Lagrange Interpolation," *SIAM Review*, 46(3) , pp. 501–517, 2004.
[21] D. A. Dunavant, "High degree efficient symmetrical Gaussian quadrature rules for the triangle," *Int. J. Numer. Meth. Eng.*, vol. 21, pp. 1129-1148, 1985.
[22] P. C. Hammer and A. H. Stroud, "Numerical Evaluation of Multiple Integrals II," *Mathematical Tables and Other Aids to Computation*, vol. 12, no. 64, pp. 272–280, 1958.
[23] L. C. Trintinalia and H. Ling, "First order triangular patch basis functions for electromagnetic scattering analysis," *J. Electromagn. Waves Appl.* , pp. 1521–1537, vol. 15, 2001.
[24] P. Ylä-Oijala, M. Taskinen, and S. Järvenpää, "Surface integral equation formulations for solving electromagnetic scattering problems with iterative methods," Radio Sci., vol. 40, RS6002, 2005.
[25] A. G. Polimeridis, J. M. Tamayo, J. M. Rius, and J. R. Mosig, "Fast and accurate computation of hypersingular integrals in Galerkin surface integral equation formulations via the direct evaluation method," *IEEE Trans. Antennas Propag.*, vol. 59, no. 6, pp. 2329-2340, June 2011.
[26] E. Ubeda, J. M. Rius, A. Heldring, and I. Sekulic, "Volumetric testing parallel to the boundary surface for a nonconforming discretization of the electric-field integral equation," *IEEE Trans. Antennas Propag.*, vol. 63, no. 7, pp. 3286-3291, July 2015.
[27] Z. Peng, X.-C. Wang, and J.-F. Lee, "Integral equation based domain decomposition method for solving electromagnetic wave scattering from non-penetrable objects," *IEEE Trans. Antennas Propag.*, vol. 59, no. 9, pp. 3328-3338, Sept. 2011.
[28] I. E. Sutherland and G. W. Hodgman, "Reentrant polygon clipping," Commun. ACM, vol. 17, pp. 32–42 , Jan. 1974.
[29] Cedric Münger, and Kristof Cools, "Global Multi-trace, Single Source Integral Equation for the Scattering by Composite Objects," *IEEE Trans. Antennas Propag.*, Early Access.
[30] Z. G. Qian and W. C. Chew, "A quantitative study on the low frequency breakdown of the EFIE," Microw. Opt. Technol. Lett., vol. 50, no. 5, pp. 1159-1162, May 2008.